# Hidden radical reactivity of the [FeO]$^{2+}$ group of the (hydro)oxide species in the H-abstraction from methane: a DFT and CASPT2 study

V. Kovalskii[a], A.A. Shubin[a,b], Y. Chen[c], D. Ovchinnikov[b], S.Ph. Ruzankin[a], J. Hasegawa[c], I. Zilberberg[a,b] and V.N. Parmon[a,b]

[a]Boreskov Institute of Catalysis, Novosibirsk 630090, Russian Federation
**E-mail:** I.L.Zilberberg@catatalysis.ru
[b]Novosibirsk State University, Novosibirsk 630090, Russian Federation
[c] Institute for Catalysis, Hokkaido University, Kita 21, Nishi 10, Kita-ku, Sapporo, Hokkaido 001-0021, Japan
**E-mail:** hasegawa@cat.hokudai.ac.jp

## Contents

**Introduction**

The participation of the radical oxygen species on the surface of metal oxides in the oxidation catalysis has been debated for decades (see e.g. review by Lunsford[1], and most recent review by Schwarz[2] and references therein). The terminal oxo center $M^{n+}$=O (where n is the oxidation state of the metal center M) is commonly considered to be the most probable candidate for this role, though there is a multiplicity of various radicaloid oxo centers in different positions.[2] The appearance of an oxo radical in the oxides implies feasibility of the electron transfer from the oxygen to the hosting metal center, e.g. as the following:

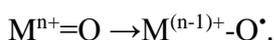
$M^{n+}$=O →$M^{(n-1)+}$-O$^•$.

Although the mentioned radical oxygen center should certainly be a quite reactive oxidant, it may relax back to "yl" oxygen state in milliseconds[3], unless some external conditions (e.g. an unusual ligand surrounding) stabilizes this radical state. This situation might be designated as an "activity-stability" dilemma.

The widely known non-photostimulated way to generate stable oxygen radicals on transition metals is attaching an oxygen atom to cations which have only a single valence electron like V(4+) or Mo(5+). The upcoming oxygen atom removes this electron from the metal to form an oxygen-centered anion radical coupled by metal cation in the maximal oxidation state like V(5+)-O$^{\bullet-}$ and Mo(6+)-O$^{\bullet-}$ (see review by Volodin et al[3]).

For oxygen on the metal centers in the intermediate oxidation state, perhaps the only experimentally identified stable radical is a species obtained by dissociation of N$_2$O on ferrous iron site in the FeZSM-5 zeolite by Panov and coworkers who named it "α-oxygen" (see review [4]). The electron configuration of this center is suggested by these authors as Fe$^{III}$-O$^{\bullet-}$ (recently they switched to designation of this state as Fe$^{III}$-O$^{\bullet}$)[5] thus assigning the oxyl radical state to the terminal oxygen on base of the reactivity data for this species. The lack of direct structural data for the α-oxygen creates some uncertainty in understanding the origin of extraordinary reactivity of α-oxygen in the C-H bond activation. In particular, for the Fe$^{III}$-O$^{\bullet}$ model, the factors are not known which stabilizes the radicaloid oxygen, i.e. prevent the oxygen in this species from taking a second electron from ferric iron to return to ferryl state Fe$^{IV}$=O. The existing quantum-chemical studies of α-oxygen seem to be still far from a consistent explanation of extraordinary reactivity of this oxidant. Consensus is not achieved yet even on nature of the ground state for the [FeO]$^{2+}$ species: whether it is oxyl Fe$^{III}$-O$^{\bullet}$ as suggested by Panov with coworkers, or ferryl Fe$^{IV}$=O. Only the latter configuration is considered as ground-state one for the [FeO]$^{2+}$ species in the coordination chemistry of dianionic oxo ligands [6], Fenton's chemistry [7], iron-oxygen active species within CuFe-ZSM-5 zeolite mediated catalytic oxidation of methane to methanol with H$_2$O$_2$ under benign conditions [8], and numerous mono-iron complexes in biomimetic chemistry. [9,10] The Fe$^{III}$-O$^{\bullet}$ assignment for α-oxygen was supported by means of the resonant inelastic X-ray scattering method which provided the evidence for the pure 3d$^5$ configuration of the iron center in the [FeO]$^{2+}$ group.[11] Quite recently, there appeared a non-radical assignment of α-oxygen to the Fe$^{IV}$=O electron structure made on the base of variable-temperature variable-field magnetic circular dichroism data in conjunction with the CASP2/B3LYP calculations.[12]

Despite the above mentioned structural uncertainty for the α-oxygen center, the designed cluster models (for instance those by Zhidomirov with coworkers [13]) allow one to make some suggestions. One of such model in which the [FeO]$^{2+}$ group is placed in the cavity of zeolite six-membered ring with two Si atoms substituted by Al atoms appeared quite useful.[13] For this model, the ferryl-type ground-state was identified by Baerends with coworkers.[14] Although being formally non-radical, the latter model possesses quite a high reactivity in the H-abstraction from methane, giving a barrier of 6.6 kcal/mol at the DFT/ZORA/BP86/TZ2P level of theory within the ADF package. Worthwhile noting that even a lower barrier of 3 kcal/mol for the same process was found by this group for a charged water-ligated complex [FeO(H$_2$O)$_5$]$^{2+}$.[15] However, for the same complex in the field of uniformly distributed counter charge, this barrier becomes 20 kcal/mol higher.[15] This seems to be a direct evidence for substantial increase of the reactivity via uncompensated charge of the ferryl group.

On the other hand, one has to mention an example of the oxyl-type ground state for the species in question modelled by simplest neutral complex FeO(OH)$_2$. For this system Malykhin showed that the ferryl-oxyl gap depends strongly on the exchange-correlation potential: for B2PLYP the ground state corresponds to ferryl configuration for the [FeO]$^{2+}$ moiety, while for M06-2X the oxyl state becomes surprisingly lower in energy than ferryl one.[16] Quite unexpectedly the next step, namely, calculation of activation energy for the H-abstraction from methane was performed only for the oxyl excited state with the B2PLYP functional. The obtained barrier of 1.5 kcal/mol appears to be quite small as it should be for a radical-like state. The M06-2X prediction of the oxyl ground state for the FeO(OH)$_2$ complex seems to contradict abovementioned BP86 predictions of the ferryl ground state for [FeO]$^{2+}$ complexes in zeolite by Baerends and our previous B3LYP results for the same complex.[17,18] Worthwhile noting that the same oxyl ground state preference for OFe(OH)$_2$ was obtained by means of the Hartree-Fock-theory based approaches: CCSD(T) [16] and CASSCF [17] which is understandable because the Hartree-Fock exchange generally "prefers" the $d^5$ maximal spin configuration for iron center in the [FeO]$^{2+}$ group.

The [FeO]$^{2+}$ species (coinciding by stoichiometry with the α-oxygen center) was suggested to appear also in the Fe(III)-hydroxides catalyzing the water-to-dioxygen oxidation.[19,20] These species are assumed to be generated by "external" water oxidizing complex Ru(bpy)$_3$$^{3+}$ via the abstraction of proton and electron from terminal hydroxo group. In particular, the iron hydroxides γ-FeO(OH) and Fe$_4$(OH)$_{10}$(SO$_4$) appear to be efficient catalysts for the water oxidation.[19] With the use of the di-iron complex Fe$_2$(OH)$_6$ and tetra-iron-hydroxide complex Fe$_4$O$_4$(OH)$_4$ which model the Fe(III) hydroxide, the authors of this paper showed that the O-O bond formation is facilitated in the oxyl type Fe$^{III}$-O$^•$ excited state.[21] The same excited-state oxyl group in the mentioned tetramer appears to abstract hydrogen from methane with a barrier as low as 5 kcal/mol while that barrier for the ground-state ferryl group is by a factor of five higher.[22]

In addition to routine computational results, some insight into reactivity of the [FeO]$^{2+}$ group has been obtained via the partition of spin density into the delocalization and polarization contributions in the basis of paired orbitals.[23] A key factor responsible for reactivity of the [Fe-O]$^{2+}$ group was shown to be the spin polarization of terminal oxygen which aligns the approaching hydrogen spin antiparallel to the oxygen spin due to Pauli exclusion principle to form the closed-shell hydroxyl anion. The majority-spin polarization of the ferryl oxygen in



Fe$^{IV}$=O results in majority-spin polarized methyl and the Fe$^{III}$ center in the excited $S = 3/2$ state in the product complex. It is in fact the high lying low-spin state for Fe$^{III}$ that makes the ferryl route unfavorable. The minority-spin polarization of the oxyl oxygen in Fe$^{III}$-O$^{\bullet}$ forces the methyl group to have the same minority-spin polarization while keeping intact the $S = 5/2$ Fe$^{III}$ state along the reaction pathway. The both described routes are presented schematically in Figure S1.

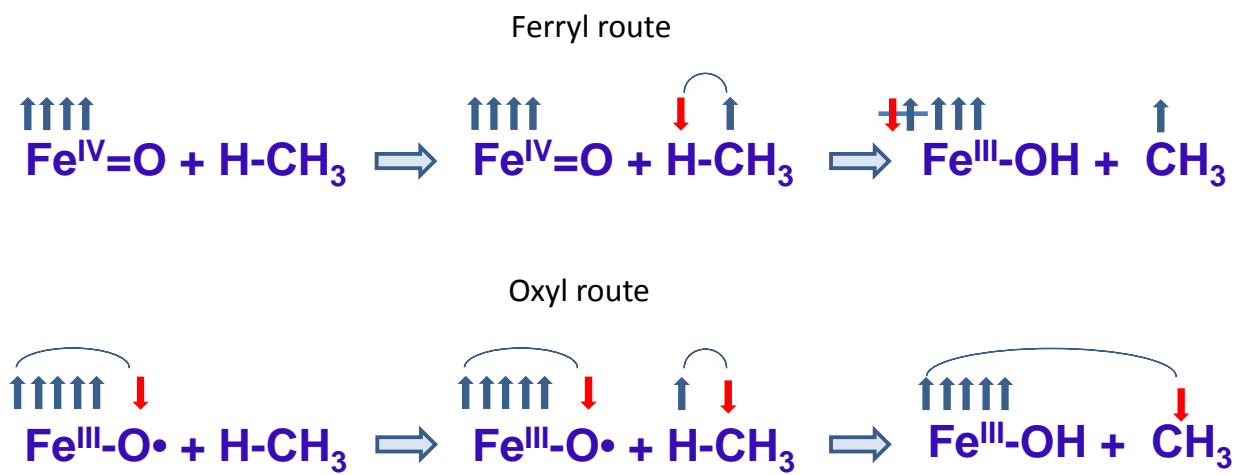

**Figure S1. Simplified scheme of the hydrogen abstraction on the ferryl and oxyl group in the iron hydroxides (modified from [22] )**

The electron configuration of the [FeO]$^{2+}$ group in various ligand surrounding is usually of the ferryl type with a negligible contribution of oxyl and only the ligand-to-metal charge transfer excited state possesses the oxyl character. Despite that, in the transition state of the H-abstraction process, the Fe$^{IV}$=O group transforms to the oxyl Fe$^{III}$-O group as was noticed in works of the Solomon's laboratory[24,25], by Ye and Neese[26], and by Dietl, Schlangen and Schwarz[2,27] on base of the DFT predictions. In the oxyl transition state, the spin density on reacting oxygen becomes negative as was pointed out by Ye and Neese[26] which reveals perhaps the most prominent feature of the oxyl oxygen. Upon the hydroxyl group formation, the β spin shifts from oxyl oxygen to the products, e.g. methyl moiety as was demonstrated for the methane H-abstraction by the [FeO]$^{2+}$ aqueous complex[15], and in our previous work for ferryl containing Fe-hydroxide tetramer complex.[22] Described data imply that the excited oxyl state is a key factor of the [FeO]$^{2+}$ group reactivity. In the course of the reaction, the oxyl term crosses the ferryl term and becomes lower lying state as was shown by Ye and Neese.[26] Moreover, one may suggest that the energy gap between ferryl and oxyl states of the [FeO]$^{2+}$ group determines the barrier of the H abstraction. This mechanism is though effective perhaps only for the ferryl-oxyl gap not larger than some threshold. If the ferryl-oxyl gap is large enough, the proton-coupled electron transfer mechanism takes place which means that only a proton of hydrogen is transferred directly to the



accepting oxygen center, while its electron goes to metal via a different route of the same complex (see Schwarz' work[27] and references therein).

In the present work the suggestion of determinative role of the oxyl state and its energy with respect to the ferryl ground state in the H abstraction from methane is proved to be true via the consideration of three simplified examples of the $[FeO]^{2+}$ group containing iron-hydroxide species and its performance in the H abstraction from methane. These model (hydr)oxide iron species are $FeO(OH)_2$, $Fe_2O(OH)_2$, and $Fe_4O_5(OH)_3$.

**Methods**

**DFT**

All DFT calculations have been performed at the UB3LYP/6-311G++(d,p) level with ultrafine integration grid within the framework of the Gaussian09 package.[28] For starting monomer the spin projection $S_z=2$ was chosen for all calculations on base of the common assumption that in ground state the $[FeO]^{2+}$ moiety possesses formally four parallel-spin $d$(Fe) electrons and closed-shell oxo dianion. In practice on this center there is always substantial positive spin density appeared apparently due to delocalization of the $d$ electrons.

For dimer and tetramer complexes (preliminary consideration of which has been made in our previous papers [17,22]) the unrestricted solutions were obtained for the maximal spin projection of the spins on iron centers) which appears to have a minimal energy among all possible iron spins configurations for the considered species. The most important DFT solutions associated with the oxyl species obtained in this work are strongly spin-contaminated and reveal negative spin density on the radicaloid oxygen center. This is not surprising since the spin contamination $\delta_S = <\hat{S}^2 - S_z(S_z+1)>$ (where spin projection $S_z = (n-m)/2$ shows the excess of the α spins over the β spins in the Kohn-Sham (KS) determinant) is determined by negative spin density ($\rho_S$) as was shown by Wang et al.[29]

In our previous works, the most reactive (oxyl) species revealed itself in spin-polarized solutions for which $\rho_S$ is negative on terminal oxygen. The sign of polarized density is defined as positive for majority spin commonly associated with the α spin, while the negative spin density corresponds to the β spin. A pairwise character of these spin-density values at the oxyl oxygen and metal allowed us to associate this configuration with a spin-polarized electron pair and designate the oxyl structure as a $^\uparrow Fe^{III}$-$O^\downarrow$ diradical.[21] It is of interest that a similar structure was earlier discussed by Yamaguchi and coworkers particularly on behalf of manganese oxygen-evolving complexes and called as metal-oxo diradical [•M-O•].[30]



## CASPT2

In the CASSCF/CASPT2 calculations, the DFT optimized geometry was used. For the basis sets, effective core potential (ECP) proposed by the Stuggart-Dresden-Bonn group [31] was used for the Fe atom. The cc-pVDZ sets were used for the others. The active space was composed of 12 electrons in 10 orbitals: five *d*-orbitals of the iron atom, three 2p orbitals of the ferryl oxygen atom, σ and σ* orbitals of the CH bond. The imaginary shift of 0.2 was used for the CASPT2 calculation for eliminating the intruder states. The calculations are performed by MOLCAS 7.8 program.[32]

## The delocalization and polarization components of the DFT solution and spin density

To get an insight into the factors responsible for the reactivity of an oxyl center, we use an approach which divides the DFT-derived spin density into the components associated with the delocalization and polarization effects.[23] Within this approach, the standard unrestricted Kohn-Sham orbitals for either spin are transformed to the biorthogonal Löwdin paired orbitals.[33] The transformation of unrestricted molecular orbitals to the paired orbitals leaves the total energy and density unchanged. The paired orbitals are naturally grouped into three mutually orthogonal subspaces of: (1) pairs of the overlapping orbitals (occupied by antiparallel spins) having the overlap integral $t$ in the $0 < t \leq 1$ range, (2) orbitals occupied by unpaired $α$ spin electrons[1], (3) unoccupied orbitals. In this basis set, one may expand the broken-spin solution in a series of restricted determinants

$$\Psi^{BS} = \sum {}^{i}C \; {}^{i}D, \qquad (1)$$

where ${}^{i}D$ are determinants built up from the occupied and unoccupied spin-up orbitals, and ${}^{i}C$ are their coefficients.[34] The ${}^{0}D$ determinant consists of only doubly and singly occupied (by spin-up electrons) orbitals, the ${}^{1}D$ determinants are generated from ${}^{0}D$ by transfer of spin-down electron from closed shells to corresponding unoccupied orbitals creating a pair of split αβ electrons for each determinant. Each ${}^{2}D$ determinant contains two split pairs simultaneously, and so on for other determinants. This series resembles the configuration interaction expansion with the singly, doubly, triply, etc. excited configurations. The weights of the determinants with the split pairs reflects the contribution of the spin-polarized configurations. Since the spin contaminants are easily annihilated from these determinants, one can estimate the contributions of pure spin-state polarized configurations into the particular BS solution.

In the paired-orbital basis set, $ρ_S$ can be divided into two components:

$$ρ_S = ρ_s^d + ρ_s^p, \qquad (2)$$

---

[1] These orbitals coincide with the unoccupied β spin orbitals in a pairwise manner



where the delocalization component $\rho_s^d$ is determined by unpaired $\alpha$ spin electrons and the polarization component $\rho_s^p$ arises from partly overlapping paired orbitals occupied by antiparallel spins.[23] It is the polarization component that is associated with the spin contamination.

**The H-abstraction on the ferryl group containing iron-hydroxide species**

## Monomer FeO(OH)$_2$

The first example is a C$_{2v}$-symmetry monomer FeO(OH)$_2$ (spin projection S$_z$=2) having the ground state corresponding to ferryl-type [FeO]$^{2+}$ moiety with a negligible oxyl contribution of 3% as estimated by annihilating septet spin contaminant from the Kohn-Sham determinant (Table S7). The paired orbitals for the ground state $^5A_1$ and the first excited state $^5B_2$ are plotted in Figure S2.



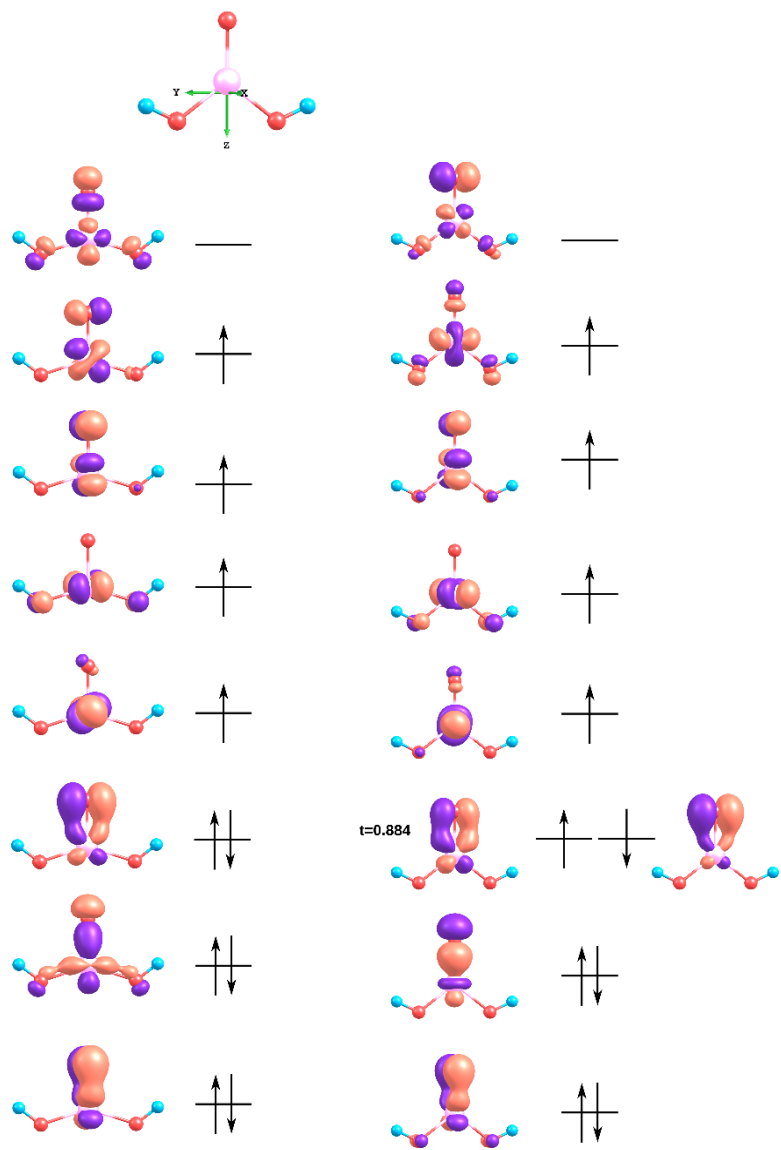

**Figure S2.** The paired orbitals localized on the [FeO]$^{2+}$ group for the C$_{2v}$-symmetry FeO(OH)$_2$ unrestricted B3LYP solutions for the ground state $^5A_1$ (left) and the excited state $^5B_2$ (right). For the split orbitals there is shown overlap integral *t*.

Despite the ferryl-type terminal oxygen in MI (having Mulliken spin density of 0.60, in Figure S3) the transition state MII reveals the negative spin density on carbon atom. This feature is a fingerprint of the oxyl route. Although the MII solution appears to be spin broken with the spin contamination of 0.68, it is almost 90% pure quintet (Table S1) indicating that the negative spin density is not an artifact associated with the septet contaminant. The spin polarization of TS results in the product complex MIII containing the β-spin methyl radical (spin density is -1.09) adsorbed on the hydroxo group formed on the terminal oxygen. The spin-polarization nature of the oxyl route for this system is highlighted by the evolution of $\rho_s^p$ on iron (0.12, 0.63, 0.77), the terminal oxygen (-0.05, -0.27, 0.05) and carbon atom (0.0, -0.50, -1.09) for the reactants, transition state and product, respectively (Table S3). Worthwhile noting that the spin density on



carbon is of pure spin-polarization nature as seen from zero delocalization contribution (Table S3).

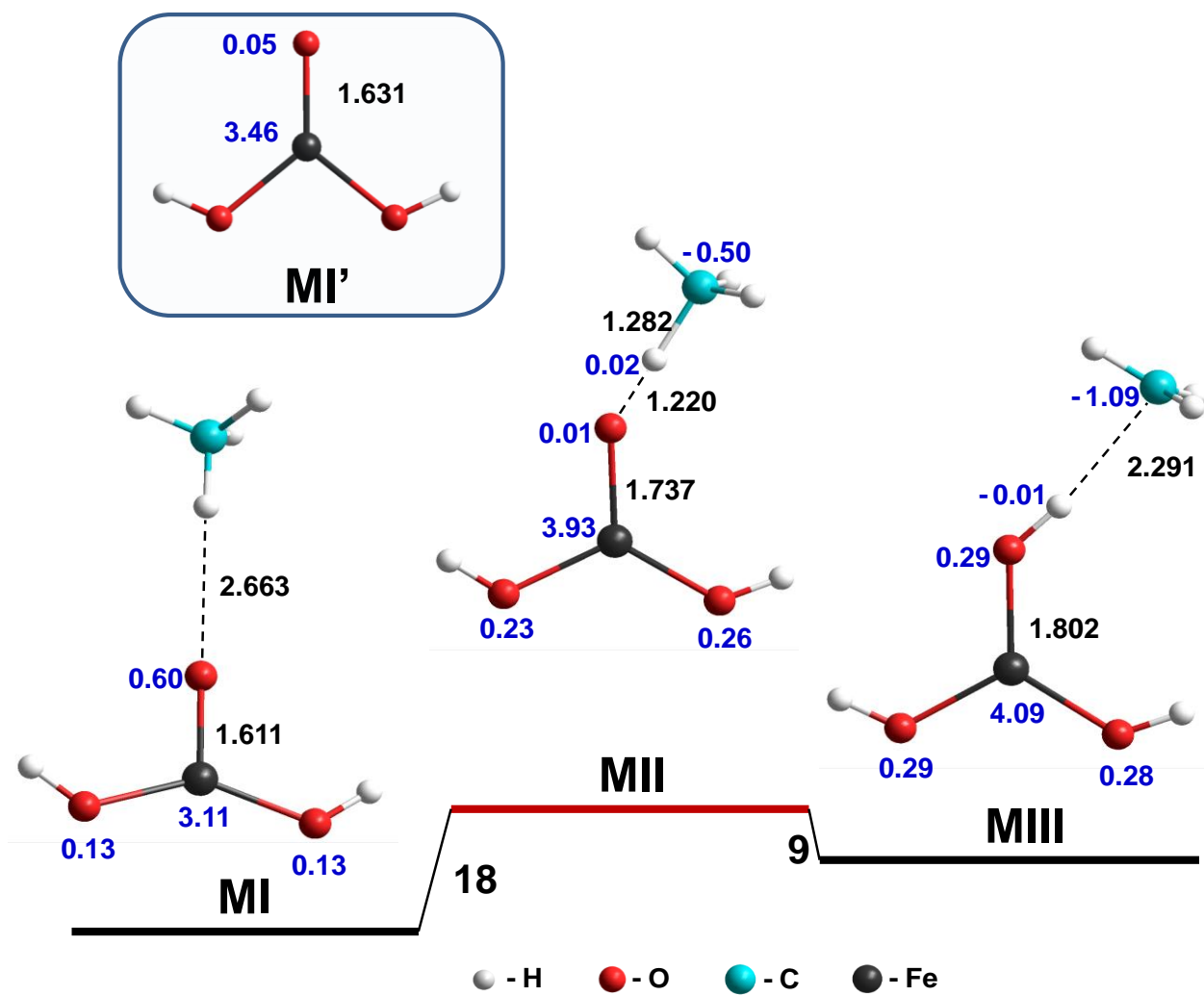

Figure S3. The hydrogen abstraction from methane at $[FeO]^{2+}$ group in monomer $O=Fe(OH)_2$ ($S_z=2$): the listed values with three and two digits after the decimal point show the distances in Å, and Mulliken spin density on atoms, respectively. Inset shows metastable oxyl-type $^5B_2$ state for the $FeO(OH)_2$ complex designated here as MI'.

The $[FeO]^{2+}$ moiety in the transition state is similar to that for the metastable oxyl species (MI' in Figure S3) lying about 15 kcal/mol higher than ferryl MI. In both cases the Fe-O bond length is somewhat elongated, the spin density on oxygen drops to almost zero, and the spin contamination is increased. The MI' structure is to be assigned to oxyl due to the polarization component of the spin density increased by magnitude from -0.05 to -0.27 for the reactants MI and transition state MII, respectively (Table S3). Without methane, the symmetry of this solution is $^5B_2$, which means that the β-spin occupied $\pi_y^*(O)$ orbital is located in the plane of molecule. Considering the transition-state structure MII having hydrogen lying in this particular plane, one may see that this oxygen orbital is certainly the key one in abstracting hydrogen from methane. There is another oxyl-type state possessing symmetry $^5B_1$ (with the energy 25 kcal/mol with



respect to ground-state ferryl) which apparently does not participate in the reaction with methane because potentially reactive $\pi_x^*(O)$ orbital is oriented perpendicular to the molecule plane yz. The MII solution is the 62% spin polarized (oxyl) state, while the solution for product complex MIII is 99% spin polarized structure (Table S9). The obtained data evidence in the necessity for the $[FeO]^{2+}$ group to transform its electron configuration from ferryl to oxyl type in order to be able removing the methane hydrogen via energy minimum oxyl route.

The competing ferryl route for considered monomer runs through a substantially higher barrier of 28 kcal/mol (Figure S4). Contrary to the oxyl case, in the ferryl-route transition state (MII-f in Figure S4) the spin density on carbon atom (0.62) is positive. The latter value increases to 1.09 for product complex MIII-f. The spin density on iron decreases upon approaching the transition state and further reflecting the lowering of the metal spin to 3/2 (as compared with S=5/2 for oxyl route).

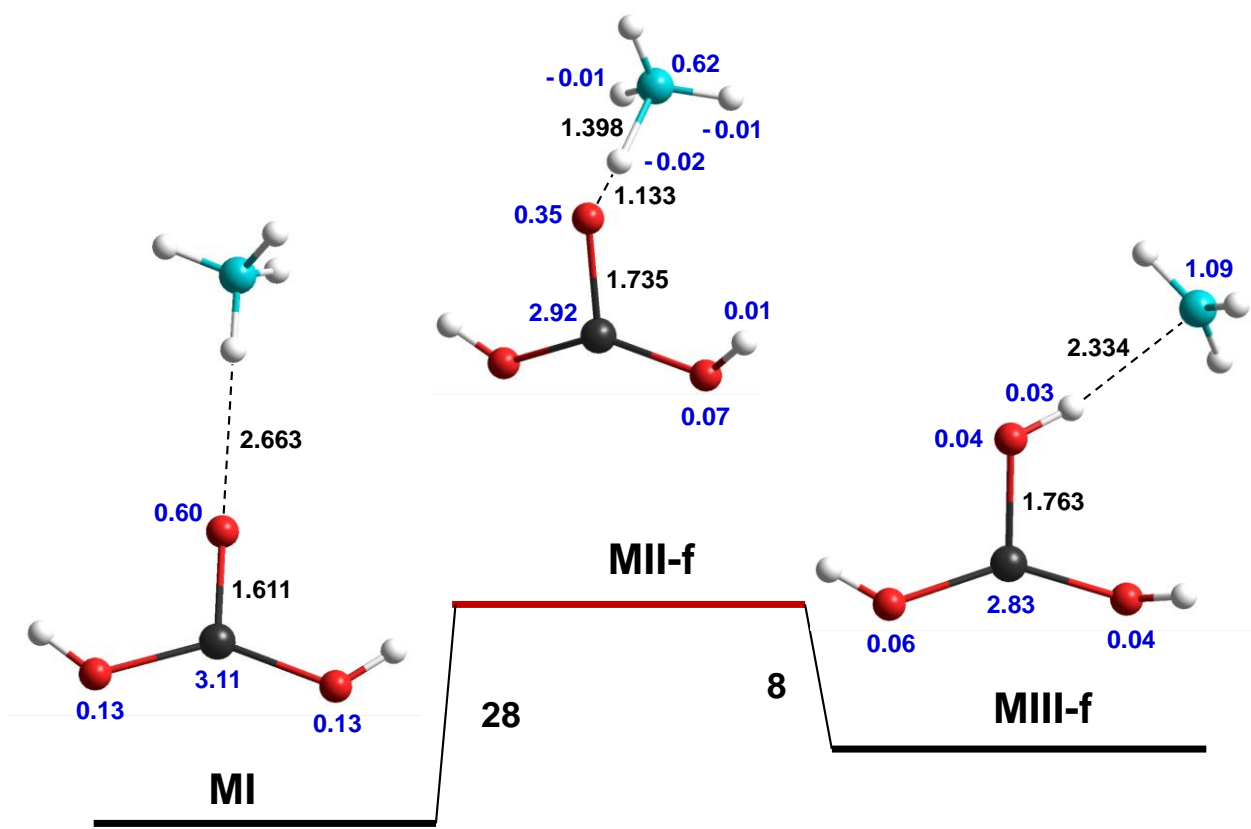

**Figure S4.** Ferryl route for the hydrogen abstraction from methane at the $[FeO]^{2+}$ group in monomer $O=Fe(OH)_2$ ($S_z=2$)

### Dimer $Fe_2O(OH)_5$

The second example is a dimer $Fe_2O(OH)_5$ with a ferryl ground state (DI) and metastable dimer state (DI') with the oxyl $[FeO]^{2+}$ group lying 13 kcal/mol above (Figure S5). This ferryl-oxyl gap is a little lower than that in case of the monomer. The H-abstraction barrier of 16 kcal/mol appears to be decreased as well with respect to the monomer case. The negative spin density of -



0.47 and -1.10 on carbon for the transition state (DII) and products (DIII), respectively, undoubtedly indicates the same oxyl-type route for the process as that found for the monomer. Transition state DII is similar to DI' as seen from the increased spin density (of about 4.0 a.u.) on iron and elongated Fe-O bond. The deviation of the spin density on terminal oxygen in DII from that in oxyl dimer DI' is apparently due to the interaction with hydrogen abstracted from $CH_4$ which causes shift of the β spin to carbon.

As in case of the monomer, the spin-polarization density reflects the appearance of a split antiparallel-spin pair on the Fe center (0.09, 0.64, 0.80), the terminal oxygen center (-0.02, -0.30, 0.05) and the carbon atom (0.0, -0.47, -1.10) for DI, DII and DIII complexes, respectively (Table S4). The β-spin electron of this pair appears finally at the $CH_3$ moiety while its α-spin counterpart remains on iron thus making it Fe(III) from Fe(IV) in the beginning of the process.

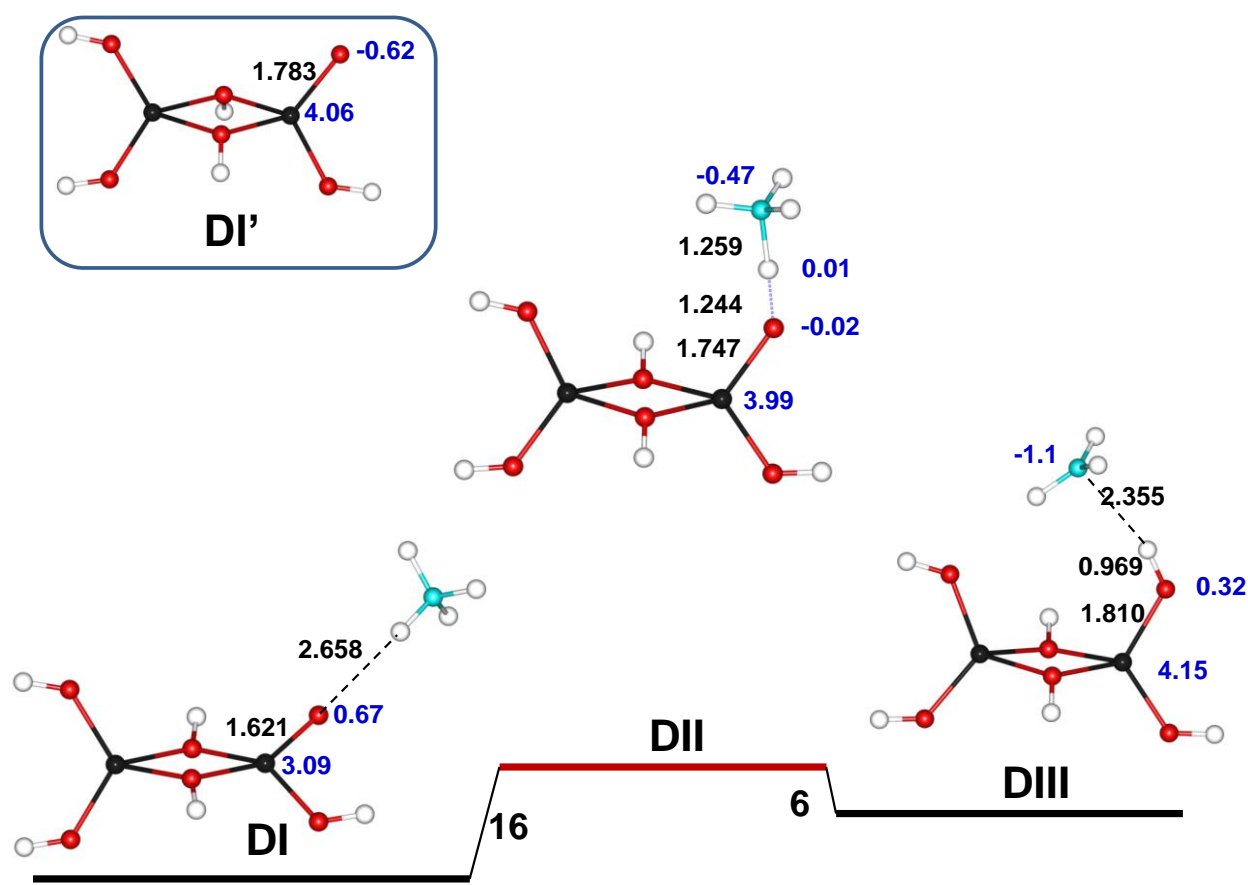

**Figure S5.** The oxyl route for the hydrogen abstraction from methane at the $[FeO]^{2+}$ group in dimer $Fe_2O(OH)_5$ with ground-state ferryl $[FeO]^{2+}$ (DI) and metastable dimer state with the oxyl group (DI')

The ferryl route for dimer species reveals about the same positive spin density on iron and carbon centers as in monomer case for transition state DII-f and the product DIII-f (Figure S6).



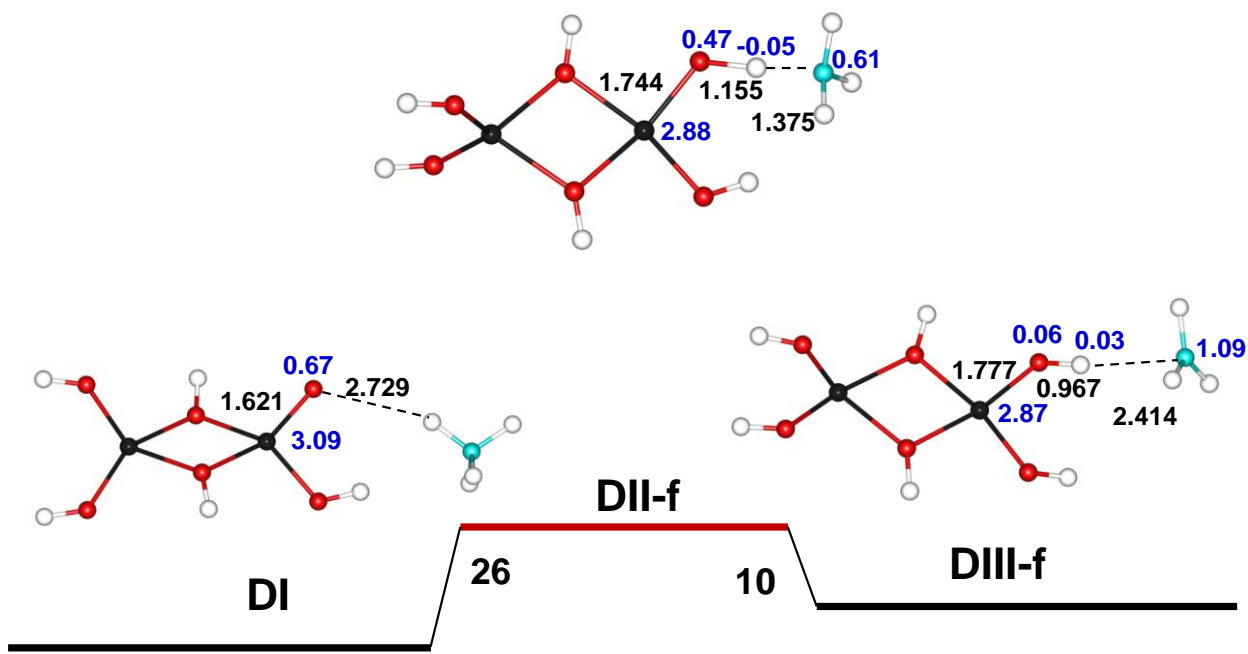

**Figure S6.** The ferryl route for the hydrogen abstraction from methane at the $[FeO]^{2+}$ group in dimer $Fe_2O(OH)_5$ with ground-state ferryl $[FeO]^{2+}$ (DI)

### Tetramer $Fe_4O_5(OH)_3$

The final system in a series of the $[FeO]^{2+}$ containing hydroxides is a tetramer $Fe_4O_5(OH)_3$ (Figure S7, Figure S8). The preliminary account of this system has been made in our previous work.[22] There have been predicted a quite low oxyl-ferryl gap of 1.4 kcal/mol for this system and distinct oxyl and ferryl routes having the barrier in the H-abstraction process of 6 and 25 kcal/mol, respectively.[22] For this system we didn't find a pathway connected the ferryl ground state to oxyl transition state, perhaps due to quasi degeneracy of ferryl and oxyl $[FeO]^{2+}$ species. Nevertheless, it seems logical to add the tetramer in the present work as a Fe-hydroxide species of the considered series with a minimal oxyl-ferryl gap, especially in the context of the correlation between the oxyl-ferryl gap and the activation energy. If one mechanistically combines the ferryl reactant (TI), and, in its turn, the oxyl transition state (TII) (Figure S7) and corresponding product complex containing β-polarized methyl radical (TIII), then this system might be considered as a limiting case for the above analyzed monomer and dimer in the sense of the lowest H-abstraction barrier. The latter value was approximated by the corresponding value for the oxyl route shifted up by the oxyl-ferryl gap of 1.4 kcal/mol. Unlike the oxyl case, the spin density appears on methyl in ferryl route (Figure S8) mostly due to the spin delocalization effect as follows from small polarization component $\rho_s^p$ for TII-f and TIII-f.



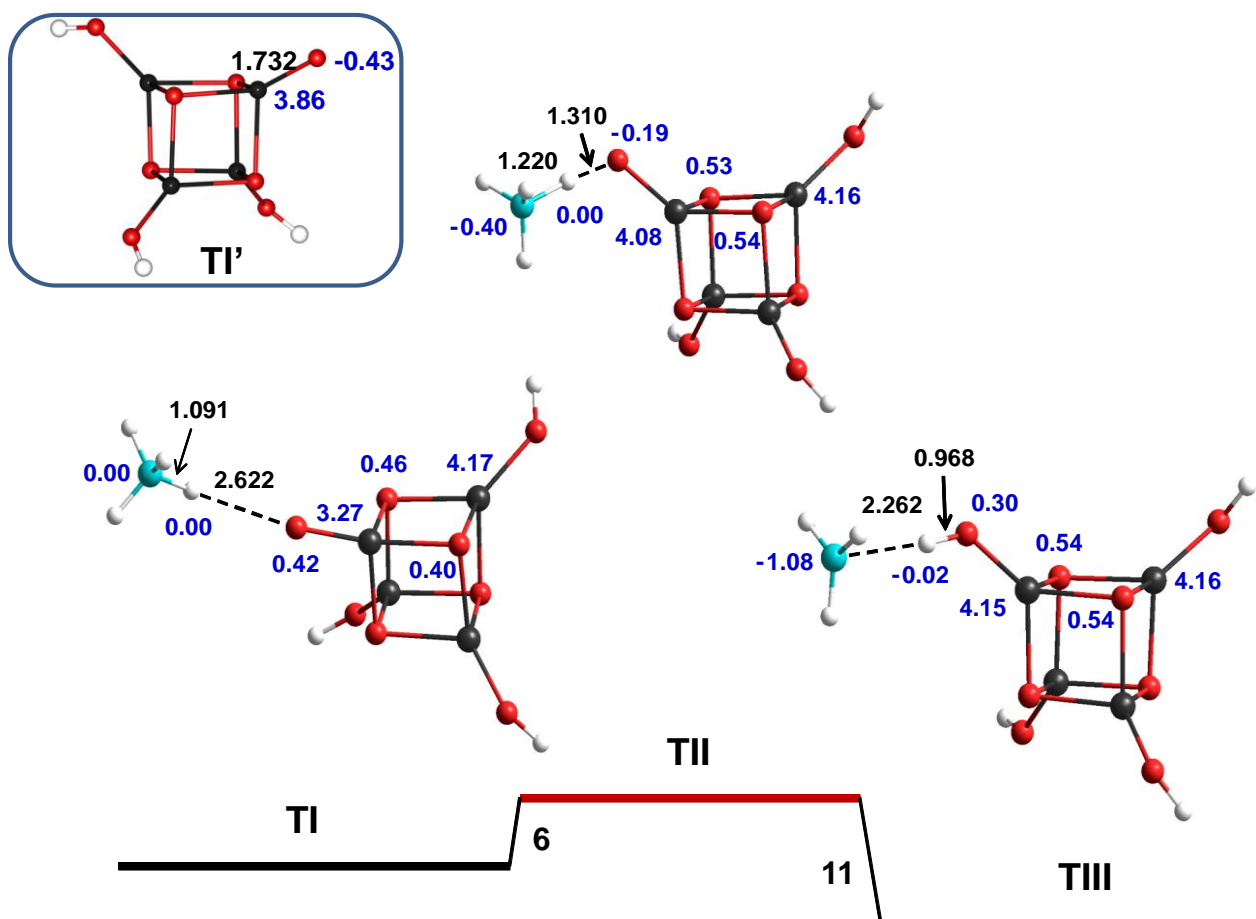

**Figure S7.** The oxyl route for the hydrogen abstraction from methane at the $[FeO]^{2+}$ group in tetramer $OFe_4O_5(OH)_3$

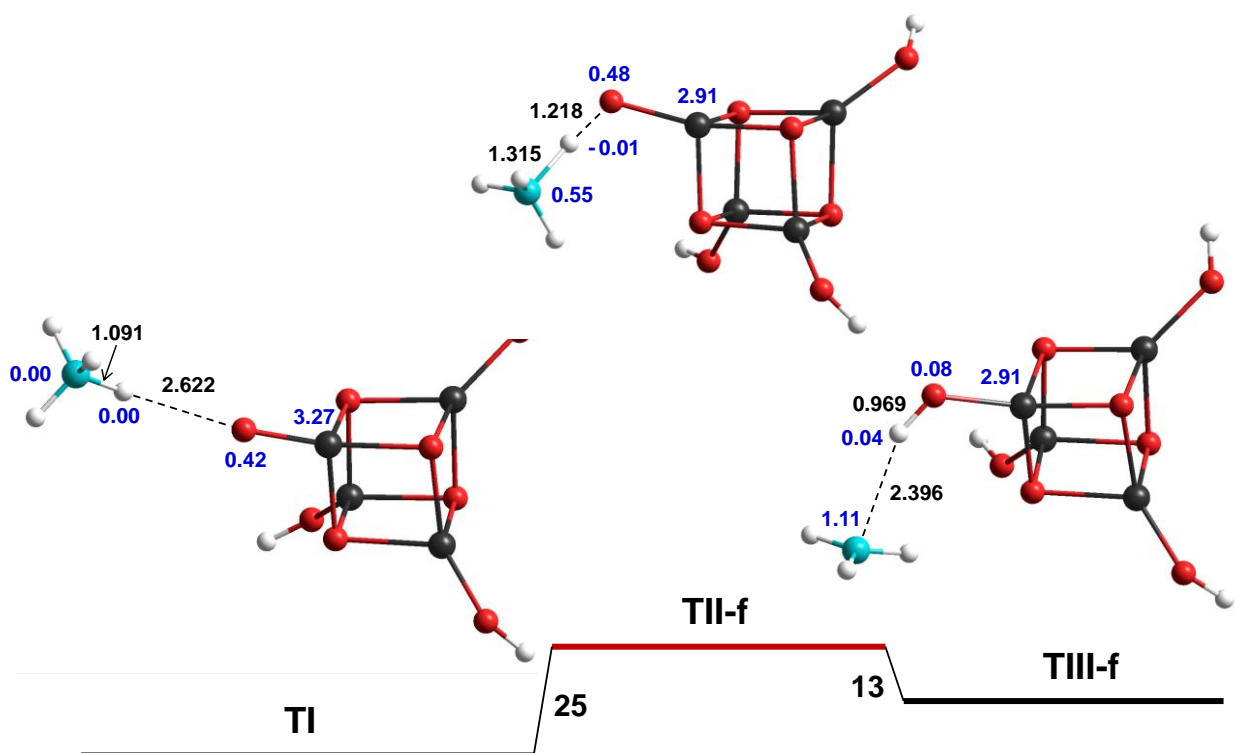

**Figure S8.** The ferryl route for the hydrogen abstraction from methane at $[FeO]^{2+}$ group in tetramer $OFe_4O_5(OH)_3$



In all three cases the spin density on iron of the [FeO]$^{2+}$ group increases by about 1 upon the OH$^-$ group formation revealing the changing of the oxidation state on iron from 4+ to 3+. As mentioned before this process is in fact responsible for the barrier of the limiting stage.

Considered above examples show that the activation energy of H-abstraction for considered species correlates with the energy gap between the ferryl and oxyl configurations for the [FeO]$^{2+}$ group (Figure S9).

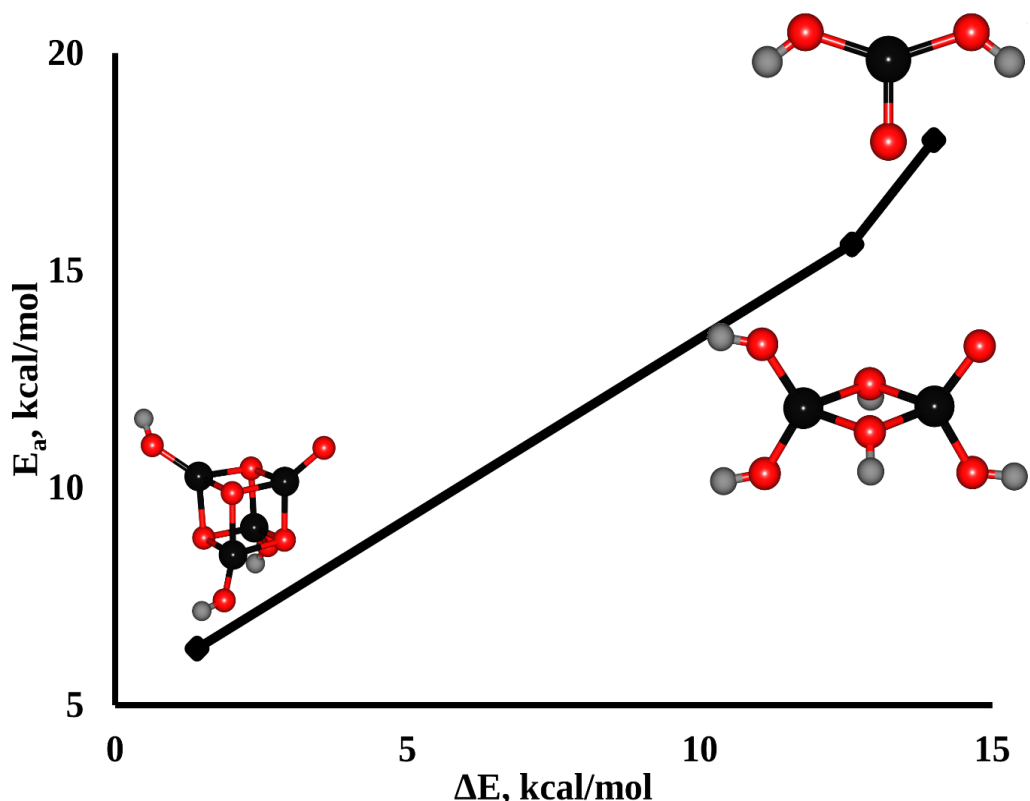

**Figure S9.** Correlation between the energy of the oxyl-ferryl gap and the activation energy of the H-abstraction from methane

The gap between the Fe$^{IV}$=O and Fe$^{III}$-O$^•$ configurations can be decreased by the ligands of the iron first coordination sphere if the electron (to be transferred from the ferryl oxygen "back" to iron) delocalizes onto these ligands in the oxyl state. This effect becomes profound for polynuclear complexes (having oxo/hydroxo bridged ligands) as was demonstrated for above considered sequence of monomer, dimer, tetramer. Worthwhile noting that the electron transferred from oxyl oxygen to iron has to be α-spin polarized to gain energy from the exchange interaction with five α spins on Fe(III) as compared to Fe(IV) in ground state. That seems to explain why the oxyl oxygen is β-spin polarized in all obtained solutions. Comparison of the polarization density on bridge-oxo centers closest to Fe for all three system with oxyl oxygen reveals relatively high values for $\rho_S^p$ for the tetramer (Table S5) thus implying the highest stabilization of the oxyl configuration and activity in the H abstraction. The found effect



highlights a strong dependence of considered process on the bridged ligands (-O- versus –OH-) in iron hydroxides.

The magnitude of polarization effect was additionally estimated using an expansion of the BS solution in a series of restricted determinants in the paired-orbital basis set (1). In all three cases under consideration the solution contains only one pair or partially overlapping orbitals implying that this expansion contains only $^0D$ and $^1D$ determinants. Since the latter determinant describes completely polarized electron pair, its weight reveals the oxyl contribution in the particular solution. So obtained oxyl weights are given in Table S1 (more detailed data are given in Table S6). One may see that for monomer and dimer the TS is more than 60% oxyl. For the tetramer, this weight reaches 74% that confirms the reactivity of this system. The 99% contribution of the spin polarization for the products in all three cases demonstrates that the unpaired electron on methyl radical is paired (correlated) with the spin-up electron on the iron center. The small weight of the contaminant (from 0.5% to maximal 17%, see Table S1) for all solutions shows that found polarization is not an artifact of the BS as often erroneously judged in literature.

**Table S1.** The monomer, dimer and tetramer of iron hydroxides: the oxyl-ferryl energy gap $\Delta E$, the barrier of the hydrogen abstraction from methane $E_a$; spin contamination $\delta_S = <\hat{S}^2 - S_z(S_z + 1)>$; the weights of the first (S+1) contaminant and polarized (oxyl) configuration for reactants, transition state, and products

| Structure | $\Delta E$, kcal/mol | $E_a$, kcal/mol | $\delta_S$ | (S+1)-contaminant weight, % | Oxyl weight, % |
|---|---|---|---|---|---|
| Monomer | 14.9 | 18 | 0.05<br>0.68<br>1.00 | 1<br>11<br>17 | 3<br>62<br>99 |
| Dimer | 12.6 | 15.6 | 0.06<br>0.70<br>1.02 | 0.5<br>6<br>9 | 3<br>65<br>99 |
| Tetramer | 1.4 | 6.3 | 0.13<br>0.80<br>1.04 | 0.5<br>4<br>5 | 8<br>74<br>99 |

**CASPT2 verification for the monomer DFT results**

Due to known shortcomings of the spin-polarized B3LYP data, especially in the case of strongly spin-contaminated states, indicating the presence of quasi-degenerate states, additional CASPT2 calculations were performed for monomer FeO(OH)$_2$ in the ferryl ground state(at the B3LYP-optimized $^5A_1$ geometry), oxyl first excited and metastable state (at the B3LYP-optimized $^5B_2$



geometry)[2]. The aim of these calculations was to confirm the B3LYP-predicted spin polarization in some states, e.g. the negative spin density on methyl moiety in the transition state and product complex for the H abstraction process. There have been calculated the structures along the reaction with methane including those of reactants, the transition-state complex with $CH_4$, and the product complex containing methyl radical using the B3LYP-optimized geometries.

The CASSCF ground state quintet for the monomer without methane reveals the same ferryl electron configuration as that predicted by B3LYP. This is clearly seen from the CASSCF natural orbitals (Figure S10) which are in fact identical to paired orbitals for the B3LYP ground-state $^5A_1$ solution (Figure S2). The CASSCF first excited state for the monomer having $^5A_1$ B3LYP geometry possesses a distinct oxyl structure as seen from the negative Mulliken spin density on terminal oxygen O1 (A in Table S2). It is worthwhile to note that the appearance of the negative spin density on O1 correlates with the increased positive spin density on iron implying that the excitation is accompanied by charge transfer from O1 to Fe. The change of spin density on Fe and O1 for the excited state with respect to the ground state (A states in Table S2) is close to the value of the polarization density estimated for the transition-state structure appeared in the process of H-abstraction from methane (Table S3). As seen from the CASSCF spin density for, the A structure in the first excited state and for the B structure (in both the ground and excited states) have the same oxyl electron structure (Table S2).

**Table S2.** The Mulliken spin densities for monomer $FeO(OH)_2$ for the CASSCF solutions obtained with the $^5A_1$ and $^5B_2$ B3LYP optimized geometry, designated as **A** and **B** structures, respectively

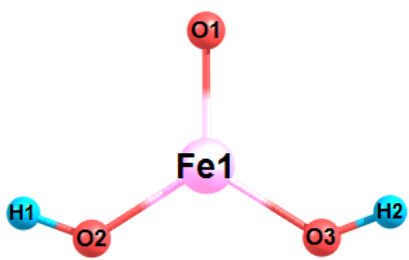

|  | Fe1 | O1 | O2 | O3 | H1 | H2 |
|---|---|---|---|---|---|---|
| **A** | | | | | | |
| Ground state | 3.72 | 0.16 | 0.05 | 0.05 | 0.00 | 0.00 |
| First excited state | 4.05 | -0.20 | 0.07 | 0.07 | 0.00 | 0.00 |
| **B** | | | | | | |
| Ground state | 4.28 | -0.44 | 0.08 | 0.08 | 0.00 | 0.00 |
| First excited state | 4.31 | -0.48 | 0.08 | 0.08 | 0.00 | 0.00 |

---

[2] In the CASSCF calculations the molecule plane was xz unlike the B3LYP data, so the $^5B_2$ state in DFT corresponds to the 5B1 state in CASSCF



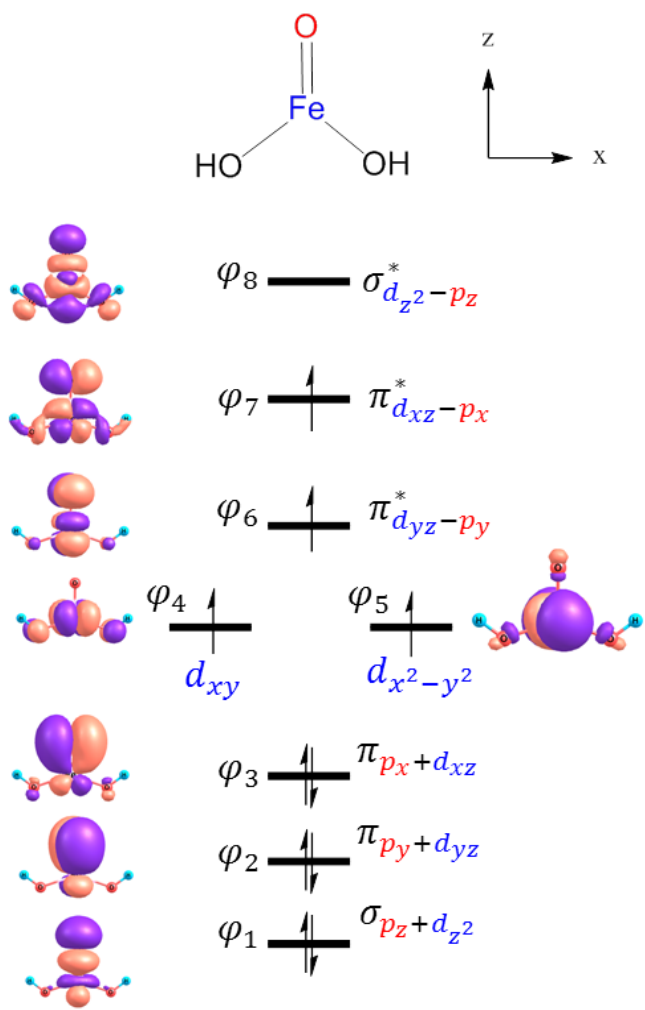

**Figure S10.** Natural orbitals along with qualitative assignments to atomic orbitals as obtained from the CASSCF solution for the S=2 ground state of the FeO(OH)2 monomer at the B3LYP geometry.

The relative energies for monomers along the oxyl H-abstraction route are given in Figure S11 (see detailed information including orbitals, relative energies and configurations in Supported materials). To distinguish between the B3LYP and CASPT2 results, the monomers MI, MI', MII, MIII are designated as R, R', TS, P, respectively. The Mulliken spin density is given for the ground and excited states of the reactants and products (Figure S11). The spin density values appear to be similar to those given by the B3LYP solution as seen from the spin density of 0.16, -0.31 and -0.64 at the terminal oxygen, and the carbon atom for the transition state (TS) and products P, respectively (Figure S11). Such a change shows the spin-polarization (oxyl) nature of this process. Moreover, the negative spin density of -0.20 and -0.44 for the first excited state of R and metastable structure R' (which might be considered as a geometry-relaxed first excited state), respectively, reveals the hidden reasons by which the transition state becomes of the oxyl type. The CASSCF wave function of the first excited state contains a 22% contribution of the configuration $\varphi_1^2 \varphi_1^2 \varphi_3^\beta \varphi_4^\alpha \varphi_5^\alpha \varphi_6^\alpha \varphi_7^\alpha \varphi_8^\alpha$ on the metal atom being in fact responsible for the oxyl



route of H-abstraction due to appearance of the β spin on the $\pi_x(O)$ orbital. It is this orbital that accepts hydrogen from methane in the FeO(OH)$_2$ plane to form spinless OH$^-$ group and adsorbed on it the β-spin polarized methyl (see all CASSCF orbitals in Supported materials).

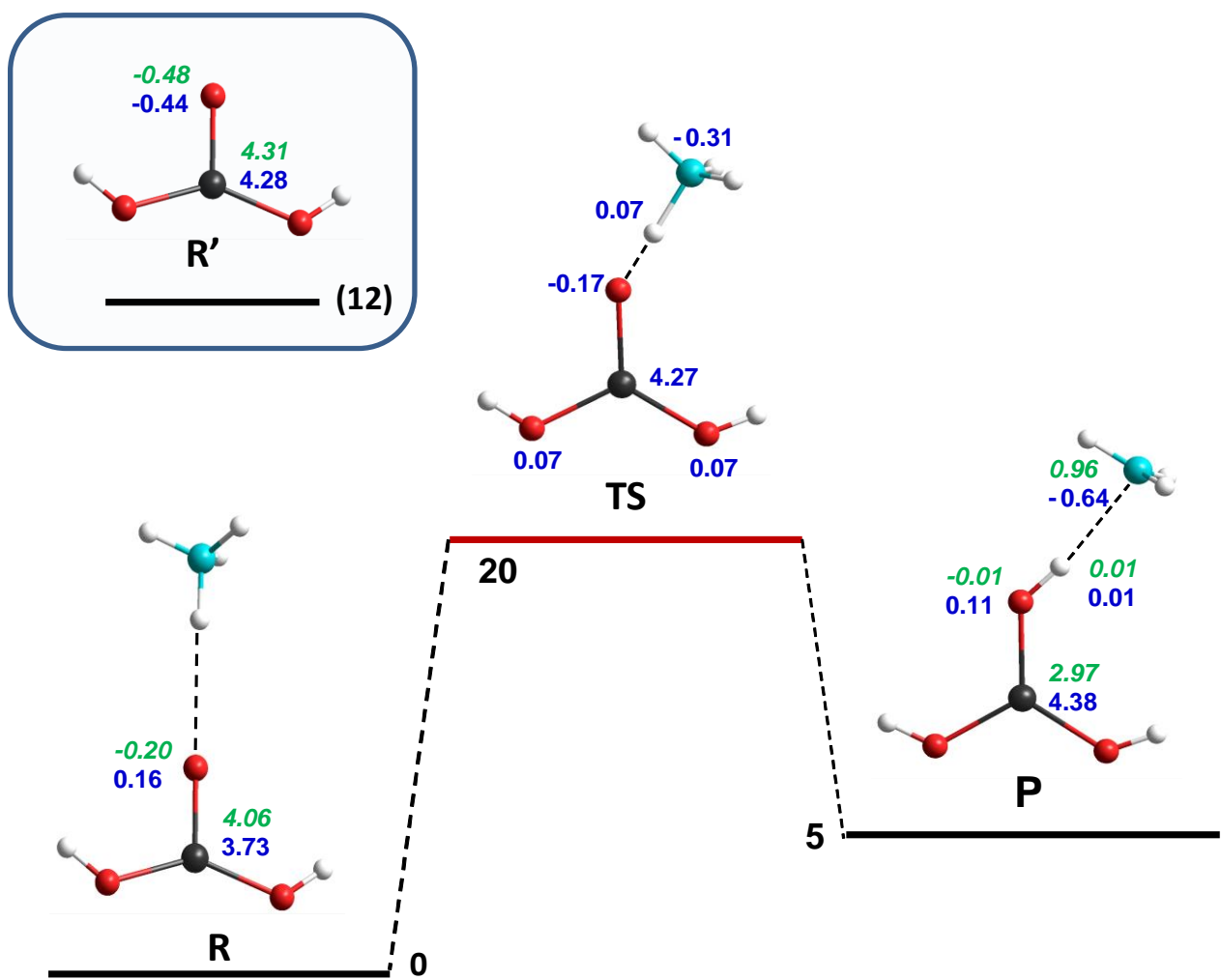

**Figure S11.** The CASSCF S=2 single-point solutions obtained for reactants R(MI), metastable state R'(MI') for the complex without methane (inset), transition state TS(MII) and products P(MIII) for the DFT-optimized geometry (see Figure S3), respectively. Relative energies are given in kcal/mol. Mulliken spin densities (a.u.) are listed for ground (lower values in blue) and first excited states (upper italic values in green).

**Discussion**

From the $\rho_s^p$ values found for the H-abstraction on the monomer, dimer and tetramer iron hydroxide species, one may see that in all cases the transition-state configuration contains a pair of spatially separated electrons with antiparallel spins at ferryl moiety: the α spin on iron center and the β spin "shared" by oxygen and carbon. This polarization is further developing in going from the transition state to products where the β spin becomes localized at carbon leaving its counter spin at iron (as the 5$^{th}$ electron in addition to its four unpaired electrons not involved in



the reaction). This evolution of the polarization density is presented in a qualitative scheme below.

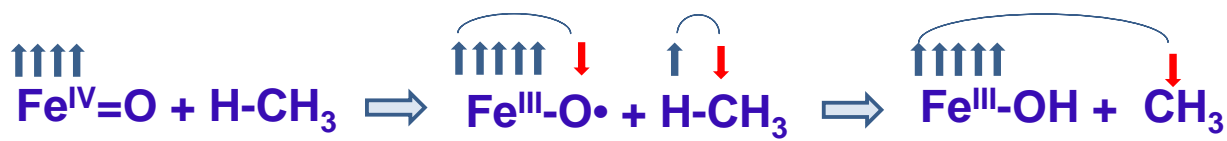

The transition state and products of the process coincide with corresponding steps of H-abstraction on the oxyl route. Since before the interaction with methane the hydrogen accepting [FeO]$^{2+}$ group possesses the ferryl ground state and oxyl as the first excited state, one must suppose that there is a ferryl-oxyl crossing somewhere before the transition state confirming the suggestion by Ye and Neese.[26] More specifically, one has to assume the avoided crossing of these quintet terms since the symmetry of the ferryl and oxyl states is the same (Figure S12). Change of the routes in the process seems to become possible because both the ferryl and oxyl configurations contribute into the ground and excited state. A small fraction of the oxyl contribution in the ground state (varying from 3 to 8%) in the beginning of the process increases up to 62-74% in the transition state thus revealing a qualitative change of the [FeO]$^{2+}$ configuration. Two-state scenario is not applicable for considered pathways since lower and higher spin states are substantially less stable than the found quintet transition state for all three systems. This agrees with other calculations for H-abstraction on the ferryl group in oxo ligand fields.[12] Quite qualitatively, one may explain the quintet preference in terms of the relative stability of the spin configurations arising from the coupling of the five α spins occupying d(Fe) orbitals, and single spin on p$_\pi$(O) resulted for ferryl group upon the Fe-O bond elongation. Competing configurations, e.g. the septet scheme with all six spins being parallel, seems far less preferable due to the Pauli Exclusion Principle which forces the oxygen spin to be antiparallel to the iron spin until the Fe-O orbital overlap is nonzero. In other words, one has almost to detach oxygen from iron to make the higher spin configuration stable. The lower spin (S<2) schemes, in their turn, require flipping spin on iron (in addition to oxygen), which is also unfavorable because of the decreased exchange interaction.



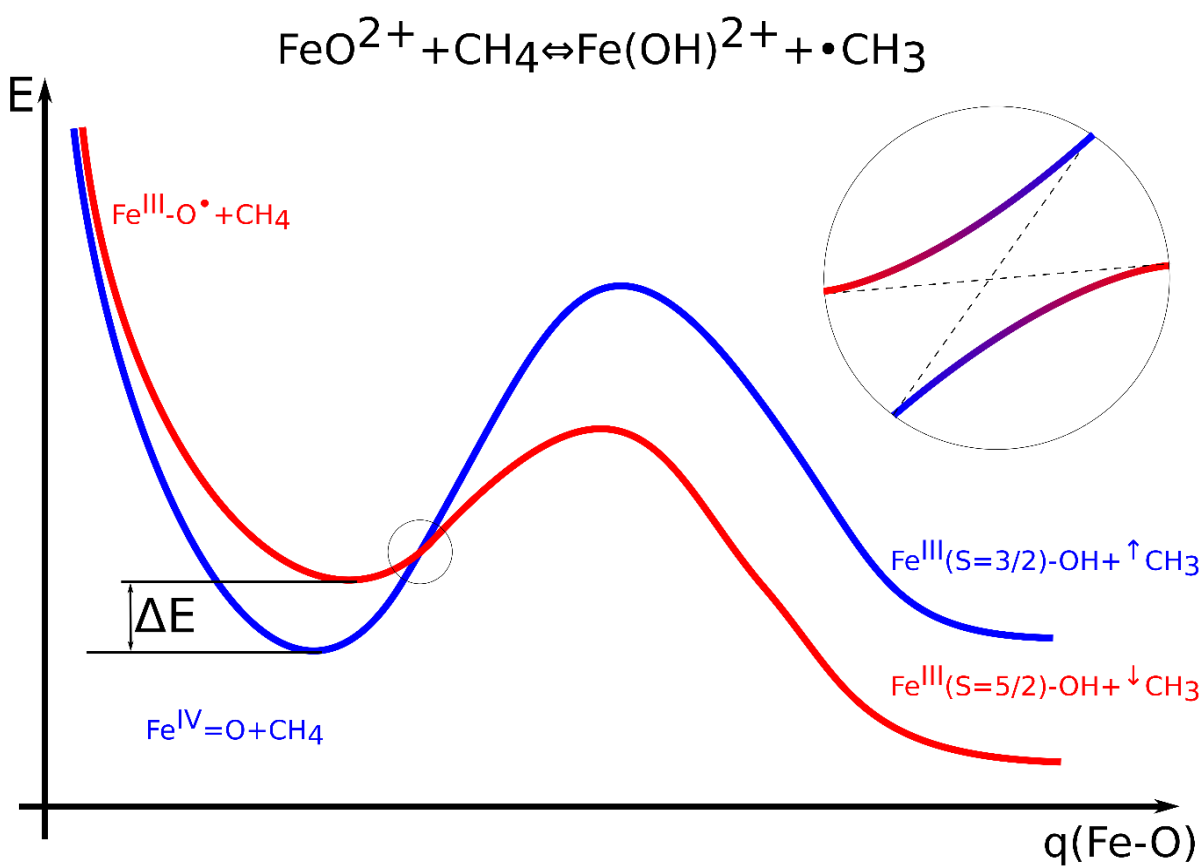

**Figure S12.** Ferryl-oxyl crossing for the reaction $FeO^{2+} + CH_4 \rightarrow Fe(OH)^{2+} + \cdot CH_3$

The negative spin density on the methyl radical appeared as a result of the H abstraction from methane by the $[FeO]^{2+}$ aqueous complex [15] indicates in fact exactly the same mechanism as that we obtained here, i.e. that the oxyl component of the $[FeO]^{2+}$ group becomes leading (within the avoided crossing mechanism) on the way to the transition state to determine the reaction pathway.

The found correlation between ferryl-oxyl gap and the activation energy resembles the correlation between the UV-visible edge energy and the activation energy for oxidehydrogenation of propane on the $V_2O_5$, $MoO_3$, $WO_3$ and $Nb_2O_5$ oxides supported by $Al_2O_3$, $ZrO_2$ and MgO.[35] Since the UV-visible edge is determined by the electron transfer from lattice oxygen to metal, this is a direct indication of the role played by radicaloid oxo centers in the hydrogen abstraction.

**Conclusion**

The H abstraction from methane by the $[FeO]^{2+}$ group in hydroxides was shown to proceed through oxyl transition state for all considered models. Therefore, the reactivity of $[FeO]^{2+}$ is determined by "hidden" first excited oxyl state $[Fe^{III}-O^\bullet]$ associated with "back" electron transfer from oxygen to iron rather than by its ferryl ground state $[Fe^{IV}=O]^{2+}$. The change of electron



configuration of the [FeO]$^{2+}$ group in the H abstraction process is explained qualitatively in terms of the avoided-crossing mechanism. One of the sequences of this mechanism is that the activation energy correlates with the energy of the ferryl-oxyl gap: the smaller gap, the lower activation energy. The minimal barrier for ferryl-containing tetramer is predicted to be as small as 6 kcal/mol, being comparable with the activation energy for the gaseous radical. These findings seem to resolve the "activity-stability" dilemma concerning the active centers in oxidation catalysis in a sense that stable non-radical (and so relatively inactive) oxygen center may possess radical-like reactivity toward H-abstraction due to the presence of low-lying excited radicaloid oxyl state. The energy of excited oxyl state with respect to ferryl state apparently depends on the particularly ligand surrounding.

As the B3LYP calculations reasonably predict, the initially excited oxyl state becomes energetically preferred in the course of reaction because this route leads to the preferable hydroxo highest-spin ferric moiety Fe$^{III}$(S=5/2)-OH in the product. The ferryl route must end at the lower-spin ferric resultant moiety Fe$^{III}$(S=3/2)-OH. The latter species is always (e.g. 17 kcal/mol for the above considered tetramer Fe-hydroxide model) higher in energy than the former due to decreased exchange interaction on iron. Thus, the spin state of ferric iron in the hydroxide product seems to determine actually the whole route of the H-abstraction.

Since the above-considered H-abstraction process takes place on a terminal oxygen alone, presented qualitative conclusions seems to be applicable to any S=2 ferryl group.

# Supplementary materials

## Monomer FeO(OH)$_2$

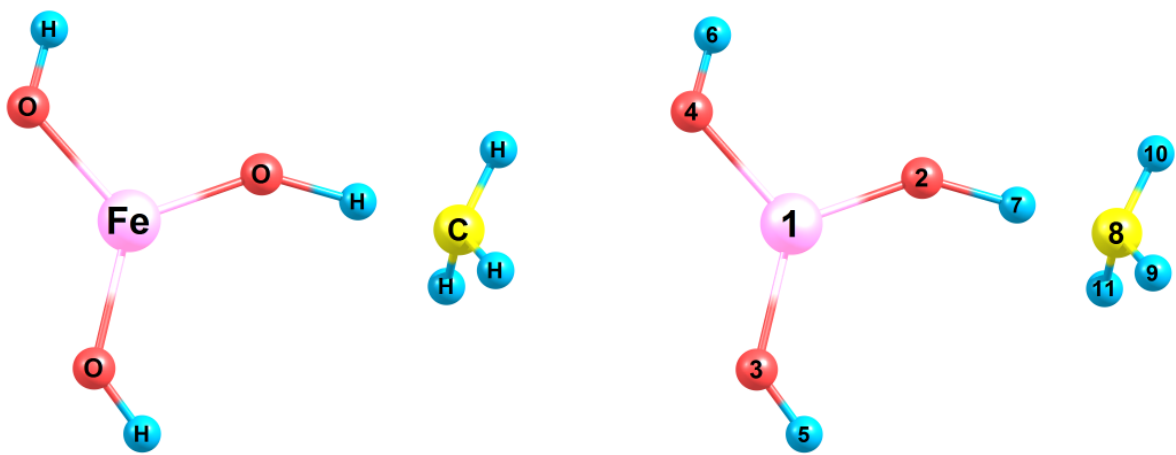

**Figure S13.** Numbering scheme for monomer FeO(OH)$_2$

**Table S3.** The spin density $\rho_S$, delocalization $\rho_S^d$ and polarization components $\rho_S^p$ for monomer FeO(OH)$_2$ in the H-abstraction from methane via the oxyl route. Numbering scheme is in **Figure S13**

|   | Atom | $\rho_S$ | | | $\rho_S^d$ | | | $\rho_S^p$ | | |
|---|------|------|------|------|------|------|------|------|------|------|
|   |      | MI   | MII  | MIII | MI   | MII  | MIII | MI   | MII  | MIII |
| 1 | Fe   | 3.11 | 3.93 | 4.09 | 2.99 | 3.30 | 3.32 | 0.12 | 0.63 | 0.77 |
| 2 | O    | 0.60 | 0.01 | 0.29 | 0.66 | 0.28 | 0.24 | -0.05 | -0.27 | 0.05 |
| 3 | O    | 0.13 | 0.26 | 0.28 | 0.16 | 0.23 | 0.24 | -0.03 | 0.04 | 0.04 |
| 4 | O    | 0.13 | 0.23 | 0.29 | 0.16 | 0.18 | 0.18 | -0.03 | 0.05 | 0.11 |
| 5 | H    | 0.01 | 0.01 | 0.01 | 0.01 | 0.01 | 0.01 | 0.00 | 0.00 | 0.01 |
| 6 | H    | 0.01 | 0.01 | 0.01 | 0.01 | 0.01 | 0.01 | 0.00 | 0.00 | 0.00 |
| 7 | H    | 0.01 | 0.02 | -0.01 | 0.00 | 0.00 | 0.00 | 0.00 | 0.01 | -0.01 |
| 8 | C    | 0.00 | -0.50 | -1.09 | 0.00 | 0.00 | 0.00 | 0.00 | -0.50 | -1.09 |
| 9 | H    | 0.00 | 0.01 | 0.04 | 0.00 | 0.00 | 0.00 | 0.00 | 0.01 | 0.04 |
| 10 | H   | 0.00 | 0.01 | 0.04 | 0.00 | 0.00 | 0.00 | 0.00 | 0.01 | 0.04 |
| 11 | H   | 0.00 | 0.01 | 0.04 | 0.00 | 0.00 | 0.01 | 0.00 | 0.01 | 0.04 |



## Dimer Fe$_2$O(OH)$_5$

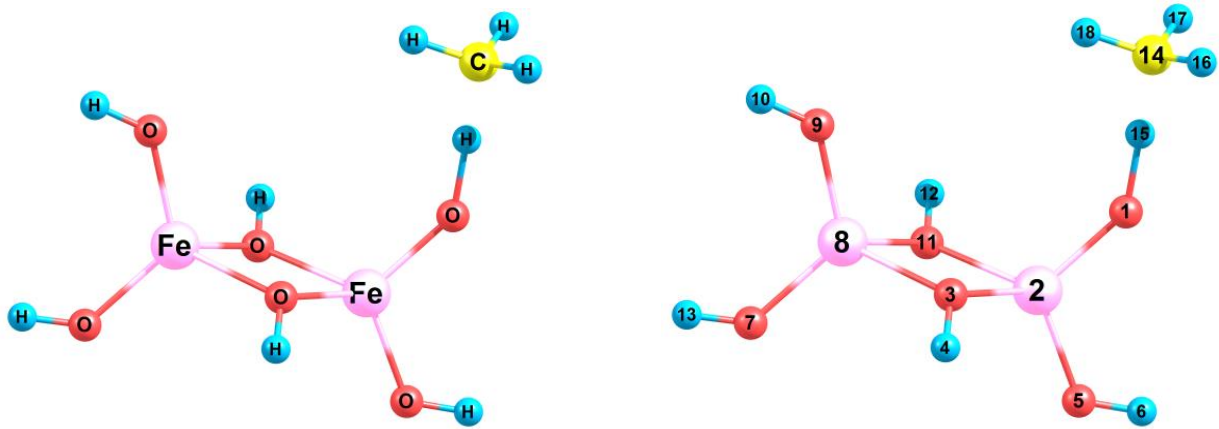

**Figure S14.** Numbering scheme for dimer Fe$_2$O(OH)$_5$

**Table S4.** The spin density $\rho_S$, delocalization $\rho_S^d$ and polarization components $\rho_S^p$ for dimer Fe$_2$O(OH)$_5$ in the H-abstraction from methane via oxyl route. Numbering scheme is in **Figure S14**

|   | Atom | $\rho_S$ | | | $\rho_S^d$ | | | $\rho_S^p$ | | |
|---|---|---|---|---|---|---|---|---|---|---|
|   |   | DI | DII | DIII | DI | DII | DIII | DI | DII | DIII |
| 1 | O | 0.67 | -0.02 | 0.32 | 0.69 | 0.28 | 0.27 | -0.02 | -0.30 | 0.05 |
| 2 | Fe | 3.09 | 3.99 | 4.15 | 3.00 | 3.35 | 3.35 | 0.09 | 0.64 | 0.80 |
| 3 | O | 0.12 | 0.19 | 0.20 | 0.13 | 0.16 | 0.17 | -0.01 | 0.03 | 0.03 |
| 4 | H | 0.01 | 0.01 | 0.01 | 0.01 | 0.01 | 0.01 | 0.00 | 0.00 | 0.00 |
| 5 | O | 0.15 | 0.26 | 0.31 | 0.19 | 0.20 | 0.23 | -0.04 | 0.06 | 0.08 |
| 6 | H | 0.01 | 0.01 | 0.01 | 0.01 | 0.01 | 0.01 | 0.00 | 0.00 | 0.00 |
| 7 | O | 0.32 | 0.31 | 0.31 | 0.29 | 0.28 | 0.28 | 0.03 | 0.03 | 0.03 |
| 8 | Fe | 4.14 | 4.14 | 4.15 | 4.19 | 4.20 | 4.18 | -0.05 | -0.06 | -0.03 |
| 9 | O | 0.32 | 0.31 | 0.30 | 0.29 | 0.28 | 0.27 | 0.03 | 0.03 | 0.03 |
| 10 | H | 0.01 | 0.01 | 0.01 | 0.02 | 0.02 | 0.02 | 0.00 | 0.00 | 0.00 |
| 11 | O | 0.14 | 0.19 | 0.19 | 0.15 | 0.18 | 0.18 | -0.01 | 0.01 | 0.01 |
| 12 | H | 0.01 | 0.01 | 0.01 | 0.01 | 0.01 | 0.01 | 0.00 | 0.00 | 0.00 |
| 13 | H | 0.01 | 0.01 | 0.01 | 0.02 | 0.02 | 0.02 | 0.00 | 0.00 | 0.00 |
| 14 | C | 0.00 | -0.47 | -1.10 | 0.00 | 0.00 | 0.00 | 0.00 | -0.47 | -1.10 |
| 15 | H | 0.00 | 0.01 | -0.01 | 0.00 | 0.00 | 0.00 | 0.00 | 0.01 | -0.01 |
| 16 | H | 0.00 | 0.01 | 0.04 | 0.00 | 0.00 | 0.00 | 0.00 | 0.01 | 0.04 |
| 17 | H | 0.00 | 0.01 | 0.04 | 0.00 | 0.00 | 0.00 | 0.00 | 0.01 | 0.04 |
| 18 | H | 0.00 | 0.01 | 0.05 | 0.00 | 0.00 | 0.00 | 0.00 | 0.01 | 0.05 |



# Tetramer Fe$_4$O$_5$(OH)$_3$

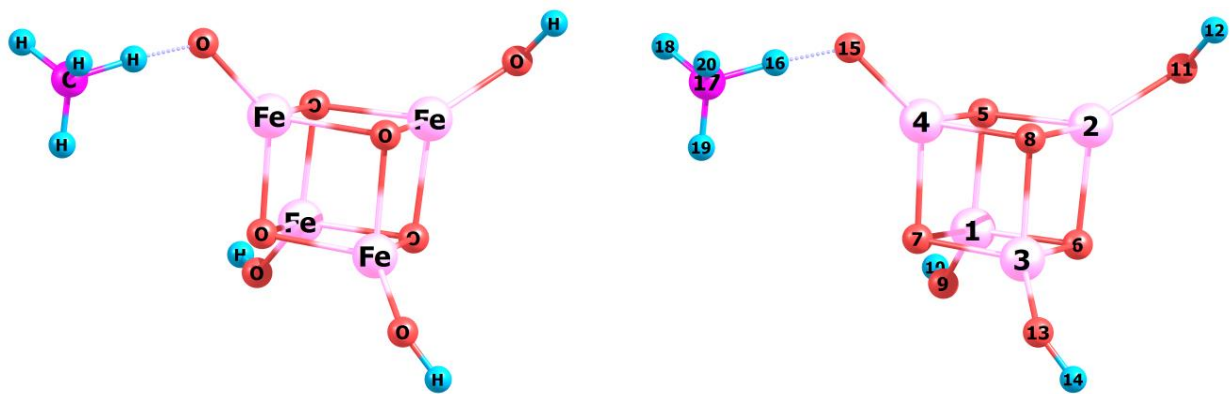

**Figure S15.** Numbering scheme for tetramer **Fe$_4$O$_5$(OH)$_3$**

**Table S5.** The spin density $\rho_S$, delocalization $\rho_S^d$ and polarization components $\rho_S^p$ for tetramer **Fe$_4$O$_5$(OH)$_3$** in the H-abstraction from methane via oxyl route. Numbering scheme is in **Figure S15**

|   |      | $\rho_S$ |      |      | $\rho_S^d$ |      |      | $\rho_S^p$ |      |      |
|---|------|------|------|------|------|------|------|-------|-------|-------|
|   | Atom | TI   | TII  | TIII | TI   | TII  | TIII | TI    | TII   | TIII  |
| 1 | Fe   | 4.16 | 4.16 | 4.16 | 4.22 | 4.22 | 4.21 | -0.06 | -0.06 | -0.05 |
| 2 | Fe   | 4.17 | 4.16 | 4.16 | 4.23 | 4.22 | 4.22 | -0.06 | -0.06 | -0.06 |
| 3 | Fe   | 4.16 | 4.16 | 4.16 | 4.23 | 4.22 | 4.21 | -0.06 | -0.06 | -0.05 |
| 4 | Fe   | 4.01 | 4.08 | 4.15 | 3.40 | 3.44 | 3.42 | 0.61  | 0.64  | 0.73  |
| 5 | O    | 0.56 | 0.53 | 0.54 | 0.51 | 0.45 | 0.46 | 0.05  | 0.08  | 0.08  |
| 6 | O    | 0.54 | 0.54 | 0.54 | 0.49 | 0.49 | 0.50 | 0.05  | 0.05  | 0.05  |
| 7 | O    | 0.50 | 0.52 | 0.54 | 0.39 | 0.40 | 0.43 | 0.11  | 0.12  | 0.11  |
| 8 | O    | 0.50 | 0.54 | 0.54 | 0.39 | 0.47 | 0.47 | 0.11  | 0.07  | 0.07  |
| 9 | O    | 0.30 | 0.29 | 0.29 | 0.27 | 0.26 | 0.26 | 0.02  | 0.02  | 0.02  |
| 10| H    | 0.01 | 0.01 | 0.01 | 0.01 | 0.01 | 0.01 | 0.00  | 0.00  | 0.00  |
| 11| O    | 0.30 | 0.29 | 0.29 | 0.27 | 0.27 | 0.26 | 0.02  | 0.02  | 0.02  |
| 12| H    | 0.01 | 0.01 | 0.01 | 0.01 | 0.01 | 0.01 | 0.00  | 0.00  | 0.00  |
| 13| O    | 0.30 | 0.29 | 0.28 | 0.28 | 0.26 | 0.26 | 0.02  | 0.02  | 0.02  |
| 14| H    | 0.01 | 0.01 | 0.01 | 0.01 | 0.01 | 0.01 | 0.00  | 0.00  | 0.00  |
| 15| O    | -0.50| -0.19| 0.30 | 0.28 | 0.25 | 0.25 | -0.78 | -0.44 | 0.05  |
| 16| H    | -0.01| 0.00 | -0.02| 0.00 | 0.00 | 0.00 | -0.01 | 0.00  | -0.03 |
| 17| C    | 0.00 | -0.40| -1.08| 0.00 | 0.00 | 0.00 | -0.00 | -0.40 | -1.08 |
| 18| H    | 0.00 | 0.01 | 0.04 | 0.00 | 0.00 | 0.00 | 0.00  | 0.01  | 0.04  |
| 19| H    | 0.00 | 0.01 | 0.04 | 0.00 | 0.00 | 0.00 | 0.00  | 0.01  | 0.04  |
| 20| H    | 0.00 | 0.01 | 0.04 | 0.00 | 0.00 | 0.00 | 0.00  | 0.01  | 0.04  |



# Ferryl and oxyl contributions in the DFT solution

The expansion of the broken-symmetry solution for considered iron hydroxide systems into a series of restricted determinants (1) revealed the leading contribution of only the $^0D$ and $^1D$ determinants describing ferryl and oxyl configuration of the $[FeO]^{2+}$ group:

$$\Psi^{BS} = {}^0C\, {}^0D + {}^0C\, {}^1D.$$

The latter determinant is a mixture of quintet and septet spin functions arising from the same configuration which can be designated as $\delta_1^\uparrow \delta_2^\uparrow \pi_1^{*\uparrow} \pi_2^{*\uparrow} \pi^\uparrow (Fe) \pi^\downarrow (O)$ skipping the closed-shell orbitals. Here the $\pi(Fe)$ and $\pi(O)$ orbitals are paired with the overlap integral which decreases in going from reactants to transition-state structures (Table S6). The absence of other contaminants follows from the fact that the $<\hat{S}^2>$ value becomes fairly close to exact value after the annihilation of septet. The weight of a single (S+1) contaminant is estimated as

$$\omega_{S+1} = \frac{\delta_S}{2(S+1)},$$

where S is the total spin, $\delta_S$ is the spin contamination[3]. Combining the weights of the $^1D$ determinant and quintet state, one can estimate the weight of oxyl configuration in the particular solution. So-obtained weights of the quintet, septet and oxyl configuration are listed in Table S6.

**Table S6.** Paired-orbital-based analysis of the DFT solutions for reactants, transition state and products for monomer, dimer and tetramer

|  | Reactants | Transition state | Products |
|---|---|---|---|
| Monomer FeO(OH)$_2$ | | | |
|  | MI | MII | MIII |
| Energy (Hartree) | -1531.19977583 | -1531.17101265 | -1531.18524820 |
| Paired-orbital overlap | 0.980 | 0.580 | 0.102 |
| C$_0$ | 0.975 | 0.576 | 0.102 |
| C$_1$ | 0.198 | 0.810 | 0.990 |
| Weights of the spin states(%): S=2, S=3 | 99.2<br>0.8 | 88.8<br>11.2 | 83.4<br>16.6 |
| Weight of spin- | 3.3 | 62.1 | 98.8 |

---

[3] I. Zilberberg, M. Ilchenko, O. Isayev, L. Gorb, J. Leszczynski, Modeling the Gas-Phase Reduction of Nitrosobenzene to Nitrosobenzene by Iron Monoxide: A Density Functional Theory Study, J. Phys. Chem. A. 108 (2004) 4878–4886. doi:10.1021/jp037351v.

| polarized structure (%) | | | |
|---|---|---|---|
| Dimer $Fe_2O(OH)_5$ | | | |
| | DI | DII | DIII |
| Energy | -3022.59736319 | -3022.57248947 | -3022.58793080 |
| Paired-orbital overlap | 0.983 | 0.571 | 0.091 |
| C0 | 0.974 | 0.568 | 0.091 |
| C1 | 0.183 | 0.817 | 0.990 |
| Weights of the spin states(%): $S=9/2$, $S=11/2$ | 99.5 0.5 | 93.8 6.2 | 90.9 0.91 |
| Weight of spin-polarized structure(%) | 3.1 | 65.3 | 0.991 |
| Tetramer $Fe_4O_5(OH)_3$ | | | |
| | TI | TII | TIII |
| Energy | -5699.39730797 | -5699.38744956 | -5699.40528241 |
| Paired-orbital overlap | 0.956 | 0.500 | 0.102 |
| C0 | 0.944 | 0.493 | 0.101 |
| C1 | 0.288 | 0.853 | 0.980 |
| Weights of the spin states(%): $S=9/2$, $S=11/2$ | 99.5 0.5 | 96.3 3.7 | 95.2 4.8 |
| Weight of spin-polarized structure (%)t | 8.1 | 74.0 | 99.0 |

**CASPT2 verification for monomer DFT results**

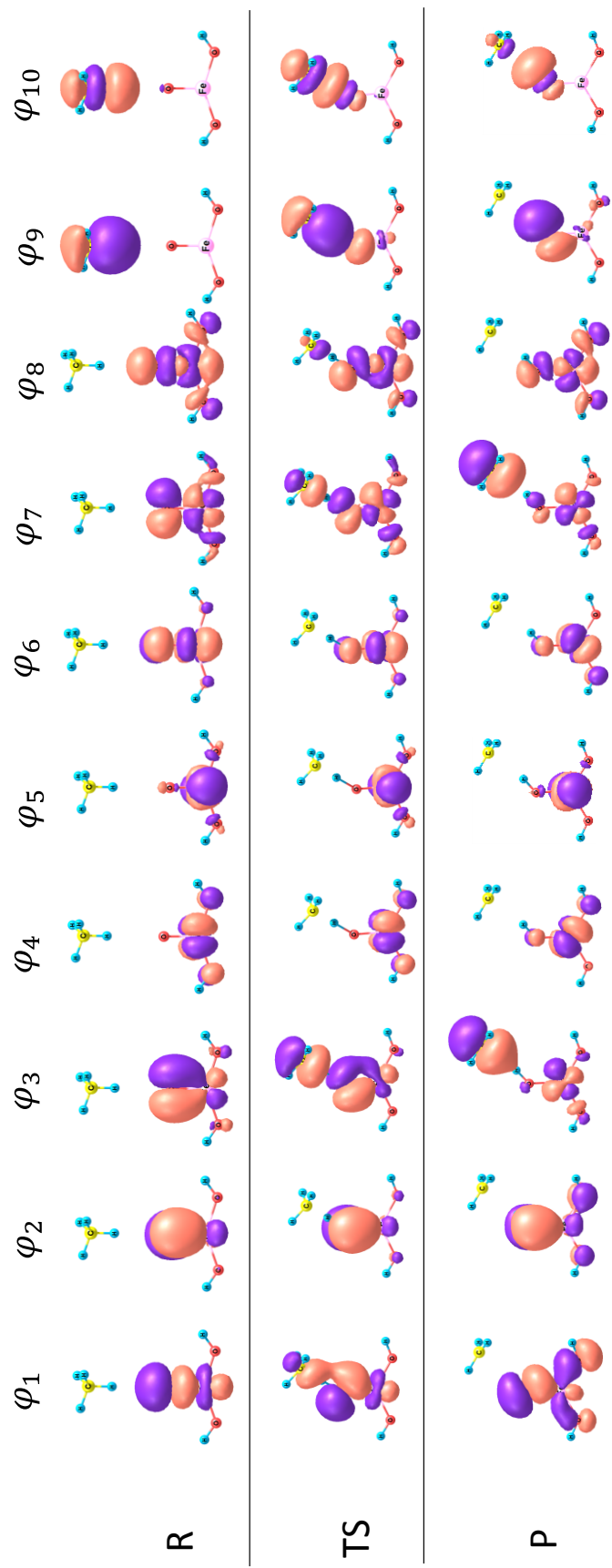

**Figure S16.** The CASSCF natural orbitals for ground-state reactants R, transition state TS and products P in the H-abstraction from methane on the monomer FeO(OH)$_2$

**Table S7.** Relative energies $\Delta E$ (in kcal/mol) and major contributions into the CASSCF wavefunction for R, TS and P in the ground and first excited state (orbitals are plotted in Figure S16)

| Structure | $\Delta E$ | Configurations | Weights, % |
|---|---|---|---|
| Ground state | | | |
| R | 0.0 | $\varphi_1^2\varphi_2^2\varphi_3^2\varphi_4^\alpha\varphi_5^\alpha\varphi_6^\alpha\varphi_7^\alpha\varphi_8^0\varphi_9^2\varphi_{10}^0$ | 72.9 |
| TS | 19.8 | $\varphi_1^2\varphi_2^2\varphi_3^2\varphi_4^\alpha\varphi_5^\alpha\varphi_6^\alpha\varphi_7^0\varphi_8^\alpha\varphi_9^2\varphi_{10}^0$ | 35.4 |
| | | $\varphi_1^2\varphi_2^2\varphi_3^\beta\varphi_4^\alpha\varphi_5^\alpha\varphi_6^\alpha\varphi_7^\alpha\varphi_8^\alpha\varphi_9^2\varphi_{10}^0$ | 33.1 |
| P | 4.5 | $\varphi_1^2\varphi_2^2\varphi_3^2\varphi_4^\alpha\varphi_5^\alpha\varphi_6^\alpha\varphi_7^0\varphi_8^\alpha\varphi_9^2\varphi_{10}^0$ | 31.1 |
| | | $\varphi_1^2\varphi_2^2\varphi_3^0\varphi_4^\alpha\varphi_5^\alpha\varphi_6^\alpha\varphi_7^2\varphi_8^\alpha\varphi_9^2\varphi_{10}^0$ | 27.5 |
| | | $\varphi_1^2\varphi_2^2\varphi_3^\beta\varphi_4^\alpha\varphi_5^\alpha\varphi_6^\alpha\varphi_7^\alpha\varphi_8^\alpha\varphi_9^2\varphi_{10}^0$ | 22.2 |
| | | $\varphi_1^2\varphi_2^2\varphi_3^\alpha\varphi_4^\alpha\varphi_5^\alpha\varphi_6^\alpha\varphi_7^\beta\varphi_8^\alpha\varphi_9^2\varphi_{10}^0$ | 15.2 |
| The first excited state | | | |
| R | 23.2 | $\varphi_1^2\varphi_2^2\varphi_3^2\varphi_4^\alpha\varphi_5^\alpha\varphi_6^\alpha\varphi_7^0\varphi_8^\alpha\varphi_9^2\varphi_{10}^0$ | 59.7 |
| | | $\varphi_1^2\varphi_2^2\varphi_3^\beta\varphi_4^\alpha\varphi_5^\alpha\varphi_6^\alpha\varphi_7^\alpha\varphi_8^\alpha\varphi_9^2\varphi_{10}^0$ | 21.6 |
| TS | 60.0 | $\varphi_1^2\varphi_2^2\varphi_3^2\varphi_4^\alpha\varphi_5^\alpha\varphi_6^\alpha\varphi_7^\alpha\varphi_8^0\varphi_9^2\varphi_{10}^0$ | 43.5 |
| | | $\varphi_1^\alpha\varphi_2^2\varphi_3^2\varphi_4^\alpha\varphi_5^\beta\varphi_6^\alpha\varphi_7^\alpha\varphi_8^\alpha\varphi_9^2\varphi_{10}^0$ | 18.3 |
| P | 51.5 | $\varphi_1^2\varphi_2^2\varphi_3^\alpha\varphi_4^\alpha\varphi_5^2\varphi_6^\alpha\varphi_7^\alpha\varphi_8^0\varphi_9^2\varphi_{10}^0$ | 87.7 |

**Table S8.** The CASSCF occupation numbers for the natural orbitals in **Figure S16**

|     | $\varphi_1$ | $\varphi_2$ | $\varphi_3$ | $\varphi_4$ | $\varphi_5$ | $\varphi_6$ | $\varphi_7$ | $\varphi_8$ | $\varphi_9$ | $\varphi_{10}$ |
|-----|------|------|------|------|------|------|------|------|------|------|
|     | Ground State(S=2) ||||||||||
| R   | 1.75 | 1.94 | 1.95 | 1.00 | 1.00 | 1.05 | 1.05 | 0.25 | 1.98 | 0.02 |
| TS  | 1.99 | 1.99 | 1.40 | 1.00 | 1.00 | 1.01 | 0.61 | 1.01 | 1.97 | 0.03 |
| P   | 2.00 | 2.00 | 1.05 | 1.00 | 1.00 | 1.00 | 0.96 | 1.00 | 1.98 | 0.02 |
|     | The First Excited State(S=2) ||||||||||
| R   | 1.98 | 1.97 | 1.54 | 1.00 | 1.00 | 1.02 | 1.02 | 0.47 | 1.98 | 0.02 |
| TS  | 1.49 | 1.98 | 1.95 | 1.00 | 1.00 | 0.51 | 1.02 | 1.04 | 1.99 | 0.02 |
| P   | 1.95 | 1.99 | 1.00 | 1.00 | 1.89 | 0.16 | 1.01 | 1.00 | 1.98 | 0.02 |

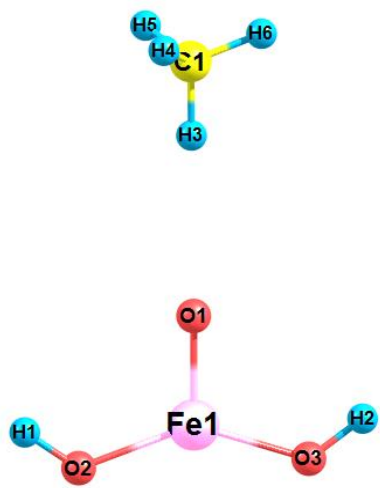

**Figure S17.** Numbering scheme for monomer FeO(OH)$_2$ in CASSCF calculations

**Table S9.** CASSCF Mulliken spin populations for R, TS and P in the ground and first excited state (numbering scheme is in **Figure S17**)

|  | Fe1 | O1 | O2 | O3 | C1 | H1 | H2 | H3 | H4 | H5 | H6 |
|---|---|---|---|---|---|---|---|---|---|---|---|
| Ground State (S=2) ||||||||||||
| R | 3.73 | 0.16 | 0.05 | 0.05 | 0.00 | 0.00 | 0.00 | 0.00 | 0.00 | 0.00 | 0.00 |
| TS | 4.27 | -0.17 | 0.07 | 0.07 | -0.31 | 0.00 | 0.00 | 0.07 | 0.00 | 0.00 | 0.00 |
| P | 4.38 | 0.11 | 0.08 | 0.07 | -0.64 | 0.00 | 0.00 | 0.01 | 0.00 | 0.00 | 0.00 |
| The First Excited State (S=2) ||||||||||||
| R | 4.06 | -0.20 | 0.07 | 0.07 | 0.00 | 0.00 | 0.00 | 0.00 | 0.00 | 0.00 | 0.00 |
| TS | 4.12 | -0.30 | 0.05 | 0.08 | 0.06 | 0.00 | 0.00 | -0.02 | 0.00 | 0.00 | 0.00 |
| P | 2.97 | -0.01 | 0.00 | 0.05 | 0.96 | 0.00 | 0.00 | 0.01 | 0.01 | 0.01 | 0.01 |

## XYZ Coordinates

### $^5A_1$ O=Fe(OH)$_2$

```
Fe   0.000000000    0.000000000    0.085335732
O    0.000000000    0.000000000   -1.514714751
O    0.000000000    1.672574325    0.613951529
H    0.000000000    2.440532679    0.037882231
O    0.000000000   -1.672574325    0.613951529
H    0.000000000   -2.440532679    0.037882231
```

**$^5B_2$ O=Fe(OH)$_2$**

| | | | |
|---|---|---|---|
| Fe | 0.000000000 | 0.000000000 | -0.103034918 |
| O | 0.000000000 | 0.000000000 | -1.733911466 |
| O | 0.000000000 | 1.410311805 | 0.963855940 |
| H | 0.000000000 | 2.286783241 | 0.564252260 |
| O | 0.000000000 | -1.410311805 | 0.963855940 |
| H | 0.000000000 | -2.286783241 | 0.564252260 |

**Monomer+CH$_4$ Reagents (Ferryl Route) (MI)**

| | | | |
|---|---|---|---|
| Fe | -0.497762000 | -1.199624000 | -0.202408000 |
| O | 0.125339000 | 0.265354000 | 0.043970000 |
| O | -2.261299000 | -1.070151000 | -0.223062000 |
| O | 0.826688000 | -2.361385000 | -0.355441000 |
| H | -2.766157000 | -0.259167000 | -0.106555000 |
| H | 1.764802000 | -2.153790000 | -0.300763000 |
| H | 1.127252000 | 2.699144000 | 0.450323000 |
| C | 1.538093000 | 3.695968000 | 0.616569000 |
| H | 1.146934000 | 4.102109000 | 1.550955000 |
| H | 1.255068000 | 4.351083000 | -0.209114000 |
| H | 2.626336000 | 3.638317000 | 0.675312000 |

**Monomer+CH$_4$ TS (Ferryl Route) (MII-f)**

| | | | |
|---|---|---|---|
| Fe | 0.304050000 | -1.228674000 | -0.199986000 |
| O | 0.809496000 | 0.276524000 | -0.899001000 |
| O | -1.385648000 | -1.034305000 | 0.288803000 |
| O | 1.734681000 | -2.266391000 | -0.108829000 |
| H | -1.970440000 | -0.312359000 | 0.042795000 |
| H | 2.579880000 | -2.110283000 | -0.538930000 |
| H | 1.225386000 | 1.123115000 | -0.270462000 |
| C | 1.750067000 | 2.224552000 | 0.412187000 |
| H | 0.937184000 | 2.514085000 | 1.071981000 |
| H | 1.957016000 | 2.915373000 | -0.399540000 |
| H | 2.625794000 | 1.812900000 | 0.905822000 |

**Monomer+CH$_4$ Products (Ferryl Route) (MIII-f)**

| | | | |
|---|---|---|---|
| Fe | -0.002516000 | -0.728163000 | -0.085941000 |
| O | 1.194393000 | 0.492667000 | -0.518047000 |
| O | -1.387075000 | 0.094571000 | 0.637464000 |
| O | 0.655168000 | -2.297717000 | -0.518548000 |
| H | -1.956297000 | 0.734075000 | 0.199348000 |
| H | 1.567830000 | -2.582141000 | -0.615764000 |
| H | 0.991164000 | 1.371815000 | -0.159081000 |
| C | 0.917466000 | 3.544178000 | 0.689927000 |
| H | 0.291428000 | 3.224230000 | 1.511740000 |
| H | 0.473908000 | 4.028351000 | -0.169284000 |
| H | 1.993691000 | 3.514927000 | 0.790820000 |

**Monomer+CH₄ Reagents (Oxyl Route) (MI)**

| | | | |
|---|---|---|---|
| Fe | -1.213134000 | -0.035496000 | -0.299897000 |
| O | 0.364174000 | 0.044548000 | 0.017623000 |
| O | -1.509415000 | 0.684085000 | -1.887875000 |
| O | -2.028147000 | -0.811515000 | 1.064222000 |
| H | -0.839771000 | 1.061238000 | -2.467091000 |
| H | -1.600952000 | -1.133130000 | 1.864358000 |
| H | 2.979604000 | 0.200921000 | 0.493435000 |
| C | 4.051501000 | 0.265677000 | 0.685647000 |
| H | 4.595207000 | -0.288491000 | -0.081366000 |
| H | 4.274175000 | -0.160453000 | 1.665372000 |
| H | 4.364430000 | 1.311025000 | 0.666810000 |

**Monomer+CH₄ TS (Oxyl Route) (MII)**

| | | | |
|---|---|---|---|
| Fe | 0.144713000 | 0.083217000 | -0.121808000 |
| O | 1.696825000 | -0.059085000 | 0.644818000 |
| O | -0.114562000 | 0.830312000 | -1.752010000 |
| O | -1.203547000 | -0.590623000 | 0.866995000 |
| H | 0.540158000 | 1.210643000 | -2.343231000 |
| H | -1.036468000 | -0.982221000 | 1.730698000 |
| H | 2.893928000 | 0.159109000 | 0.556235000 |
| C | 4.152924000 | 0.385587000 | 0.469021000 |
| H | 4.472359000 | -0.175525000 | -0.405862000 |
| H | 4.545862000 | 0.005089000 | 1.408608000 |
| H | 4.230738000 | 1.464581000 | 0.359330000 |

**Monomer+CH₄ Products (Oxyl Route) (MIII)**

| | | | |
|---|---|---|---|
| Fe | -0.684127000 | -0.103334000 | 0.005183000 |
| O | 0.883761000 | -0.289511000 | 0.873872000 |
| O | -0.852491000 | 0.653708000 | -1.633288000 |
| O | -2.200586000 | -0.702212000 | 0.797000000 |
| H | -0.151586000 | 1.016955000 | -2.181329000 |
| H | -2.217826000 | -1.118670000 | 1.664290000 |
| H | 1.789187000 | -0.043671000 | 0.636456000 |
| C | 4.025065000 | 0.398055000 | 0.400035000 |
| H | 4.186902000 | -0.194160000 | -0.490244000 |
| H | 4.118790000 | -0.062374000 | 1.374018000 |
| H | 3.941652000 | 1.473420000 | 0.321695000 |

**Dimer+CH₄ Reagents (Ferryl Route) (DI)**

| | | | |
|---|---|---|---|
| O | -1.669576129 | -1.045055078 | 0.182384014 |
| Fe | -0.778752062 | 0.307619023 | 0.124514010 |
| O | 0.544147044 | 1.536358116 | 0.910607067 |
| H | 0.300235023 | 2.452174187 | 1.075512081 |
| O | -2.063207160 | 1.531817119 | -0.036231003 |
| H | -2.958338227 | 1.173763087 | -0.067981005 |
| O | 2.986686225 | 1.965860152 | -1.016259078 |
| Fe | 2.229300171 | 0.815540062 | 0.159279012 |
| O | 3.300099251 | -0.074828006 | 1.318548099 |
| H | 4.253581323 | -0.044974003 | 1.429821110 |
| O | 0.850027067 | -0.422172032 | -0.606053045 |
| H | 0.909424070 | -1.370698104 | -0.754063059 |
| H | 3.909671299 | 2.181420166 | -1.170968089 |
| C | -3.538136270 | -4.288866327 | 0.384660029 |
| H | -3.006216232 | -3.338277257 | 0.325716025 |
| H | -4.451100340 | -4.161932316 | 0.969078073 |
| H | -3.796871289 | -4.627697356 | -0.620090049 |
| H | -2.903341222 | -5.035190386 | 0.865584069 |

**Dimer+CH$_4$ TS (Ferryl Route) (DII)**

| | | | |
|---|---|---|---|
| O | -2.880505440 | -1.384998865 | -0.355466047 |
| Fe | -1.845734271 | 0.013995834 | -0.510126327 |
| O | -0.512611166 | 0.422284333 | 0.926380110 |
| H | -0.559884129 | 1.236194027 | 1.437270735 |
| O | -2.795199443 | 1.477280054 | -0.992723464 |
| H | -3.747554912 | 1.436841715 | -1.124755077 |
| O | 2.230775856 | 1.361546455 | -0.203488718 |
| Fe | 1.148972784 | -0.086550529 | -0.046021567 |
| O | 1.884038798 | -1.680526700 | 0.433871479 |
| H | 2.792751432 | -1.874500172 | 0.676896597 |
| O | -0.158333469 | -0.412617147 | -1.509202618 |
| H | -0.087909915 | -1.170193141 | -2.097074202 |
| H | 3.183209014 | 1.437180145 | -0.108278526 |
| C | -2.740031571 | -3.688316085 | 0.613709938 |
| H | -2.825928771 | -2.531028969 | 0.124727875 |
| H | -3.423033180 | -3.671818153 | 1.459874107 |
| H | -3.055033298 | -4.346390355 | -0.192882739 |
| H | -1.694817471 | -3.799249755 | 0.892931393 |

**Dimer+CH$_4$ Products (Ferryl Route) (DIII)**

| | | | |
|---|---|---|---|
| O | -2.684019207 | -0.855615068 | -0.331381025 |
| Fe | -1.550522117 | 0.550954041 | -0.446899034 |
| O | -0.272931021 | 0.757563059 | 1.076444084 |
| H | -0.238662018 | 1.570088118 | 1.590913121 |
| O | -2.269495173 | 2.146747166 | -0.932380070 |
| H | -3.201355242 | 2.293117175 | -1.119136088 |
| O | 2.700952207 | 1.220032091 | 0.218709017 |
| Fe | 1.353035104 | 0.009680001 | 0.209010016 |
| O | 1.698187128 | -1.722160133 | 0.671219054 |
| H | 2.531025196 | -2.093582162 | 0.974036072 |
| O | 0.144848011 | -0.018783001 | -1.367329103 |
| H | 0.156144012 | -0.736452056 | -2.007308151 |
| H | 3.635926278 | 1.118894085 | 0.412166031 |
| C | -1.373681103 | -3.560737271 | 1.029884079 |
| H | -2.471971190 | -1.704327130 | 0.085153006 |
| H | -1.955891152 | -3.482940267 | 1.938584150 |
| H | -1.698993129 | -4.239345322 | 0.252407019 |
| H | -0.403110031 | -3.084061237 | 0.965216074 |

**Dimer+CH4 Reagents (Oxyl Route) (DI)**

| | | | |
|---|---|---|---|
| Fe | -0.972315511 | -0.250716999 | 0.058962077 |
| O | 0.549670993 | 0.038422778 | 1.306354961 |
| H | 0.486705044 | 0.659622769 | 2.038357277 |
| O | -1.971884465 | 1.249730757 | -0.025101884 |
| H | -1.602227978 | 2.085657722 | -0.330495474 |
| O | 2.705298239 | 1.778029947 | -0.058323145 |
| Fe | 2.044746531 | 0.088787624 | -0.014152750 |
| O | 3.177197475 | -1.322944250 | -0.033189149 |
| H | 4.136515948 | -1.364749030 | -0.027692122 |
| O | 0.520749606 | -0.261881298 | -1.258546215 |
| H | 0.573700265 | -0.935239707 | -1.944357384 |
| H | 3.611235450 | 2.087018146 | -0.137484773 |
| O | -1.826036500 | -1.777349743 | 0.138297758 |
| C | -5.188431068 | -0.882619978 | 0.030069706 |
| H | -4.449958090 | -1.682995381 | 0.104478154 |
| H | -5.861456194 | -0.928574900 | 0.888145652 |
| H | -4.677034085 | 0.080992766 | 0.017496818 |
| H | -5.764435778 | -1.003006405 | -0.889237514 |

**Dimer+CH4 TS (Oxyl Route) (DII)**

| | | | |
|---|---|---|---|
| Fe | -1.837391628 | -0.054188874 | -0.007341311 |
| O | -0.309704565 | -0.018314708 | 1.288988560 |
| H | -0.282780466 | 0.585695210 | 2.036624290 |
| O | -2.715302288 | 1.540482545 | -0.053794836 |
| H | -2.345179585 | 2.389084480 | -0.313746421 |
| O | 2.132105937 | 1.343001941 | -0.031443790 |
| Fe | 1.189024961 | -0.209815070 | -0.000073061 |
| O | 2.086471332 | -1.785608358 | -0.007693681 |
| H | 3.029929691 | -1.963913839 | 0.001880989 |
| O | -0.336604714 | -0.317207225 | -1.287693732 |
| H | -0.373311093 | -1.027859193 | -1.935893427 |
| H | 3.078533441 | 1.480592502 | -0.119048350 |
| O | -2.861763686 | -1.502318414 | 0.048924796 |
| C | -5.373759191 | -1.265697357 | 0.035123980 |
| H | -4.161321825 | -1.433754188 | 0.043100891 |
| H | -5.723440999 | -1.744924254 | 0.947465109 |
| H | -5.526425505 | -0.189134465 | 0.024553832 |
| H | -5.717053777 | -1.761088184 | -0.871003854 |

**Dimer+CH4 Products (Oxyl Route) (DIII)**

| | | | |
|---|---|---|---|
| Fe | -1.079807222 | -0.162523885 | -0.013004300 |
| O | 0.429957894 | 0.048610702 | 1.287463836 |
| H | 0.391135194 | 0.630643531 | 2.051590132 |
| O | -2.156764401 | 1.307125554 | -0.057572853 |
| H | -1.901660681 | 2.198616426 | -0.312246537 |
| O | 2.687670721 | 1.711001822 | -0.014212704 |
| Fe | 1.945304675 | 0.052195589 | 0.000582463 |
| O | 3.034711754 | -1.397137646 | -0.013861346 |
| H | 3.993041687 | -1.455419234 | 0.000117717 |
| O | 0.447595766 | -0.230762929 | -1.291644257 |
| H | 0.501126039 | -0.924953835 | -1.956160646 |
| H | 3.610130834 | 1.965237163 | -0.096594029 |
| O | -1.994993905 | -1.725350347 | 0.024971516 |
| C | -5.165079180 | -0.681980005 | 0.060735493 |
| H | -2.961849196 | -1.751800118 | 0.013738184 |
| H | -5.496098780 | -1.073321382 | 1.013237962 |
| H | -4.517469645 | 0.184678591 | 0.024201382 |
| H | -5.575491401 | -1.092475006 | -0.851993421 |

**Tetramer+CH4 Reagents (Ferryl route) (TI)**

| | | | |
|---|---|---|---|
| Fe | 0.935683819 | -1.565242923 | 0.826471779 |
| Fe | 1.542617156 | 0.344518319 | -1.231585973 |
| Fe | 1.363984432 | 1.182477499 | 1.502899578 |
| Fe | -0.969440845 | 0.383334821 | 0.083328985 |
| O | 0.117111563 | -0.940863110 | -0.874136801 |
| O | 2.328892848 | -0.187045750 | 0.509498460 |
| O | -0.086953699 | -0.166843368 | 1.732166536 |
| O | 0.447773609 | 1.636944953 | -0.194544207 |
| O | 1.282779940 | -3.261963828 | 1.282949562 |
| H | 0.819361944 | -3.910194756 | 1.818684127 |
| O | 2.459559803 | 0.699324716 | -2.730611401 |
| H | 2.441828719 | 0.347670994 | -3.623642756 |
| O | 1.891448666 | 2.324866667 | 2.778519416 |
| H | 2.730860162 | 2.520026183 | 3.201206538 |
| O | -2.491110726 | -0.162106115 | 0.124058766 |
| H | -5.095861695 | -0.366224377 | -0.099374039 |
| C | -6.179616077 | -0.375656900 | -0.224414078 |
| H | -6.650781423 | -0.713361275 | 0.700192445 |
| H | -6.530522030 | 0.629994252 | -0.462594076 |

| H | -6.447820146 | -1.054107639 | -1.036059842 |

**Tetramer+CH4 TS (Ferryl route) (TII-f)**
| Fe | 0.135188046 | -1.363563699 | 0.542286658 |
| Fe | 1.372433192 | 0.275031580 | -1.436584090 |
| Fe | 0.211252226 | 1.480985512 | 0.890435291 |
| Fe | -1.456924997 | 0.337279951 | -1.058412602 |
| O | -0.025371018 | -1.073888015 | -1.369076388 |
| O | 1.507519501 | 0.072330993 | 0.541233690 |
| O | -1.193958301 | 0.081221613 | 0.815721416 |
| O | -0.033241621 | 1.597733104 | -1.066719319 |
| O | 0.241528307 | -2.877742907 | 1.504534709 |
| H | 0.106131136 | -3.800187537 | 1.275973069 |
| O | 2.763045395 | 0.479986604 | -2.558178958 |
| H | 3.134146641 | -0.070357281 | -3.251237263 |
| O | 0.224028138 | 2.855676094 | 2.049188196 |
| H | 0.871552484 | 3.162992379 | 2.687563219 |
| O | -3.007353042 | -0.312829815 | -1.442671856 |
| H | -3.868914951 | -0.249649553 | -0.583678833 |
| C | -4.743902631 | -0.138495375 | 0.391176796 |
| H | -5.336637354 | -1.043369362 | 0.277839721 |
| H | -4.132679673 | -0.105991962 | 1.289281226 |
| H | -5.284305900 | 0.780167680 | 0.176283294 |

**Tetramer+CH4 Products (Ferryl route) (TIII-f)**
| Fe | 0.077498793 | -1.439598273 | 0.659885516 |
| Fe | 1.626430217 | 0.075129276 | -1.171608484 |
| Fe | 0.335046318 | 1.394935322 | 1.040636024 |
| Fe | -1.236008048 | 0.340686601 | -1.057866660 |
| O | 0.211612218 | -1.203485714 | -1.242047040 |
| O | 1.553451732 | -0.086446453 | 0.811584653 |
| O | -1.144400909 | 0.108797698 | 0.837869395 |
| O | 0.222669163 | 1.480324786 | -0.948190823 |
| O | -0.054125000 | -2.931920778 | 1.658080827 |
| H | 0.041532705 | -3.858489794 | 1.425765231 |
| O | 3.132650525 | 0.266206961 | -2.137473763 |
| H | 3.498087485 | -0.229621121 | -2.873536128 |
| O | 0.354693111 | 2.798194342 | 2.162892660 |
| H | 0.989106966 | 3.095172784 | 2.819326819 |
| O | -2.799018102 | -0.204609491 | -1.625737842 |
| H | -3.578406150 | -0.325586915 | -1.062944682 |
| C | -4.710636354 | -0.492191397 | 1.041832701 |
| H | -5.070550587 | -1.506450992 | 0.932690920 |
| H | -3.700764594 | -0.310709325 | 1.386867517 |
| H | -5.407114450 | 0.330487528 | 0.950587980 |

**Tetramer+CH4 Reagents (Oxyl route) (TI)**

| | | | |
|---|---|---|---|
| Fe | 0.619487040 | -0.923541154 | -0.768448679 |
| Fe | 1.120078394 | 1.532881485 | 0.587388281 |
| Fe | -0.281083257 | -0.580347837 | 1.935595839 |
| Fe | -1.561399196 | 0.840726200 | -0.161815827 |
| O | 0.263236508 | 0.966945944 | -1.065379347 |
| O | 1.400534022 | -0.394685992 | 0.979877724 |
| O | -1.128721218 | -0.996804395 | 0.179209517 |
| O | -0.656826507 | 1.320864022 | 1.458154948 |
| O | 1.281501426 | -2.194471175 | -1.845167157 |
| H | 1.436882096 | -2.261762809 | -2.789914973 |
| O | 2.303960170 | 2.834458960 | 0.930281062 |
| H | 2.658832328 | 3.570813933 | 0.427226790 |
| O | -0.692593576 | -1.356418854 | 3.496376612 |
| H | -0.190292319 | -1.819527636 | 4.170701894 |
| O | -2.952961812 | 1.534591447 | -0.929970393 |
| H | -3.883517417 | 0.006928509 | -2.747572707 |
| C | -3.845359890 | -0.984650511 | -3.202997009 |
| H | -3.239481024 | -0.946623532 | -4.109914542 |
| H | -3.409472262 | -1.697286035 | -2.500552471 |
| H | -4.857315661 | -1.303657120 | -3.457033379 |

**Tetramer+CH4 TS (Oxyl route) (TII)**

| | | | |
|---|---|---|---|
| Fe | 1.342399483 | -0.856685059 | -0.110279226 |
| Fe | 1.084141215 | 1.730690640 | 1.096735496 |
| Fe | -0.273550529 | -0.506035722 | 2.224273474 |
| Fe | -1.121332380 | 0.553657676 | -0.297768549 |
| O | 0.783260790 | 0.921244805 | -0.690491811 |
| O | 1.561357153 | -0.095593575 | 1.710317681 |
| O | -0.547081570 | -1.185682805 | 0.383266482 |
| O | -0.763366355 | 1.283942701 | 1.513977750 |
| O | 2.432173303 | -2.056141681 | -0.886436055 |
| H | 3.034503465 | -2.011942796 | -1.632359161 |
| O | 2.045937767 | 3.212084718 | 1.426469089 |
| H | 1.820781516 | 4.078135373 | 1.773552378 |
| O | -0.945246481 | -1.236613014 | 3.722304716 |
| H | -0.543838999 | -1.608798169 | 4.510650884 |
| O | -2.444510272 | 1.081043406 | -1.330468870 |
| H | -3.437305133 | 0.339566091 | -1.756961263 |
| C | -4.326214245 | -0.417996634 | -2.108113178 |
| H | -4.339382371 | -0.336361889 | -3.193210400 |
| H | -4.050327717 | -1.409377394 | -1.754816492 |
| H | -5.224412703 | -0.026971553 | -1.634388725 |

**Tetramer+CH4 Products (Oxyl route) (TIII)**

| | | | |
|---|---|---|---|
| Fe | 1.432113860 | -1.044334651 | -0.627692458 |
| Fe | 1.130010897 | 1.548800222 | 0.550343588 |
| Fe | -0.196598689 | -0.693446705 | 1.699961714 |
| Fe | -1.058225739 | 0.325393479 | -0.837056107 |
| O | 0.851570794 | 0.715713372 | -1.227901597 |
| O | 1.633486801 | -0.263027434 | 1.186816980 |
| O | -0.455324248 | -1.396357607 | -0.135958079 |
| O | -0.712214426 | 1.077957531 | 0.971152168 |
| O | 2.541576471 | -2.239608602 | -1.383573595 |
| H | 3.137765155 | -2.197457396 | -2.134523208 |
| O | 2.064703282 | 3.050799600 | 0.867756579 |
| H | 1.827473449 | 3.907517404 | 1.229538708 |
| O | -0.858581286 | -1.422357530 | 3.203843688 |
| H | -0.449003185 | -1.806758033 | 3.982055417 |
| O | -2.435022902 | 0.745458865 | -1.898638555 |
| H | -3.200769745 | 0.262046629 | -2.239887722 |
| C | -5.048590572 | -0.735230122 | -3.082365171 |
| H | -4.775240741 | -0.428196850 | -4.082675878 |
| H | -4.821041122 | -1.738266563 | -2.747990133 |
| H | -5.694074996 | -0.104655375 | -2.486314290 |